# A critical perspective for emerging ultra-thin solar cells with ultra-high power-per-weight outputs

Apostolos Panagiotopoulos,[1,#] Temur Maksudov,[2,#] George Kakavelakis,[3,4,#,*] George Perrakis,[5] Essa A. Alharbi,[6] Dimitar Kutsarov,[1] Furkan H. Isikgor,[2] Salman Alfihed,[6] Konstantinos Petridis,[4] Maria Kafesaki,[5,7] S. Ravi P. Silva,[1] Thomas D. Anthopoulos,[2,*] Michael Graetzel[3]

## Affiliations

[1] Advanced Technology Institute (ATI), Department of Electrical and Electronic Engineering, University of Surrey, Guildford, Surrey GU2 7XH, UK.

[2] KAUST Solar Center (KSC), Physical Sciences and Engineering Division (PSE), King Abdullah University of Science and Technology (KAUST), Thuwal, 23955-6900 Saudi Arabia.

[3] Laboratory of Photonics and Interfaces, Institute of Chemical Sciences and Engineering, Ecole Polytechnique Federal de Lausanne, 1015, Lausanne, Switzerland.

[4] Department of Electronic Engineering, School of Engineering, Hellenic Mediterranean University, Romanou 3, Chalepa, GR-73100, Chania, Crete, Greece.

[5] Institute of Electronic Structure and Laser (IESL), Foundation for Research and Technology - Hellas (FORTH), 70013 Heraklion, Crete, Greece

[6] Microelectronics and Semiconductors Institute, King Abdulaziz City for Science and Technology (KACST), Riyadh 11442, Saudi Arabia

[7] Department of Materials Science and Technology, University of Crete, 70013 Heraklion, Crete, Greece

[#] These authors contributed equally

[*] Authors to whom correspondence should be addressed: georgios.kakavelakis@epfl.ch; kakavelakis@hmu.gr; and thomas.anthopoulos@kaust.edu.sa

# Abstract

Ultrathin, solution-processed emerging solar cells with high power-per-weight (PPW) outputs demonstrate unique potential for applications where low weight, high power output, and flexibility are indispensable. The following perspective explores the literature of emerging PVs and highlights the maximum reported PPW values of Perovskite Solar Cells (PSCs) 29.4 W/g, Organic Solar Cells (OSCs) 32.07 W/g and Quantum Dot Solar Cells (QDSC) 15.02 W/g, respectively. The record PPW values of OSCs and PSCs are approximately one order of magnitude higher compared to their inorganic ultrathin solar cells counterparts (approx. 3.2 W/g for CIGS and a-Si). This consists emerging PVs, very attractive for a variety of applications where the PPW is the key parameter. In particular, both OSCs and PSCs can be implemented in different scenarios of applications (indoor and biocompatible applications for OSCs and outdoor and high-energy radiation conversion conditions for the PSCs) due to their unique optoelectronic and physiochemical properties. Finally, our theoretical optical and electrical simulation and optimization study for the most promising and well-suited PV technologies, showed an impressive maximum realistic theoretical PPW limit of 74.3 and 93.7 W/g for PSCs and OSCs, respectively. Our finding shows that the literature PSCs and OSCs towards high PPW outputs, is not quite close to the theoretical maximum and thus more work needs to be done to further increase the PPW output of these promising PV technologies.

# I. Introduction

The world's population reached eight billion in 2022 and is expected to continue growing with approximately a 1% growth rate per annum, and thus will reach 9.7 billion people in 2050.[1] Since global energy demand will likely continue to grow by 39 percent annually by 2030, more energy sources should be deployed to cover global energy demands.[2] Unquestionably, over the last decades, fossil fuels (such as coal, oil, and natural gas) formed the world's primary energy source by supplying almost 80% of total demand.[3,4] However, while fossil fuels are burned to generate energy to power transportation and industrial processes, greenhouse gases (GHG) are released into earth's atmosphere.[5] The increased emitted GHG due to anthropogenic activities and their accumulation are the primary cause of climate change, altering the Earth's ecosystems.[6] In 2015, to tackle the negative impacts, governments at the UN Climate Change Conference (COP21) reached a breakthrough, the Paris Agreement,[7] under which, governments set a pathway towards a Net-Zero Emissions world for around 2050.[8–10]

To fulfil the promises for a Net-Zero emissions policy by 2050, renewable energy sources have to be significantly exploited, as they have the lowest carbon footprint among other energy generation technologies. Among renewables, photovoltaic (PV) technology has one of the most significant climate change mitigation potentials, with a 20 times lower carbon footprint of approximately 50 g $CO_2$ per·kW·h$^{-1}$ (data correspond to silicon solar panels) compared to 1000 g $CO_2$ per·kW·h$^{-1}$ for fossil fuels (natural gas).[11] Within the last decade, PVs contributed more than half of all renewable power expansion, followed by wind and hydropower.[12] Based on the latest energy overviews for 2021, solar PV reported an increment of 151 gigawatts of renewable power addition (the vast majority of them, i.e. ~95%, based on silicon solar panels), remaining the most powerful renewable technology.[12] By the end of 2023, solar PV led renewable additions, reaching another milestone and pushing through the 300 GW mark. However, to meet the Net Zero Emissions scenario by 2050, solar PV technology requires a 25% annual growth by 2030, corresponding to a more than threefold increase.

Although silicon-based PVs have dominated the solar panels market up to date (more than 95% market capture), challenges such as the high cost of their production, low throughput, the lack of high specific power (the power-per-weight ratio), flexibility and portability stand along the path.[13] Multiple markets are appealing for incorporating the latter added value propositions in applications like aerospace, electric vehicles, building integrated PVs, wearables, and self-powered microscale IoT indoor and outdoor-integrated PV products.[14] As a result, integrating rigid solar cells into applications that require lightweight devices remains a significant bottleneck; thus, flexible ultra-lightweight solar cells can, in principle, overcome this complication.[15]

Unquestionably, the interest in emerging PV technologies[16], including Perovskite Solar Cells (PSCs), Organic Solar Cells (OSCs) and Quantum Dot Solar Cells (QDSC), remains unwavering.[17,18] During the last decade, solution-processed semiconductors, such as metal halide perovskites, fullerene, non-fullerene-based organic molecules, polymeric semiconductors and novel inorganic quantum dots have been significantly developed. Particularly, PSCs have demonstrated remarkable power conversion efficiencies (PCE) up to 26% on rigid substrates, competing with state-of-the-art silicon technology.[16] Inversely, OSCs with the rise of non-fullerene molecule acceptors, have surpassed 19% of PCE moving on the road to 20%.[16] While on the other hand QDSCs have demonstrated PCE values reaching above 18%.[16] The key advantage of emerging PVs, embrace cost-effective, solution-processed materials with low-temperature processing, compatible with printing fabrication techniques to produce suitable flexible, high PPW solar cells.[19]

Due to the high absorption coefficient values of emerging thin film materials (due to their direct band gap), the active layer's thickness can significantly be reduced ranging from 100 to 500 nm, i.e. three orders of magnitude thinner than silicon or poly-Si absorbers where their thickness is around 100 to 200 μm. Thus, metal halide perovskites, organic molecules and quantum dots can form solar cells in thin films and further minimize the total weight of the devices for lightweight, flexible applications. An additional advantage of flexible solar cells that contributes to the reduction of the weight and cost compared to the glass substrates, is the replacement of rigid substrates with ultrathin, light, flexible substrates such as polyethylene terephthalate (PET)[20], polyethylene naphthalate (PEN)[20], polydimethylsiloxane (PDMS)[21], cellulose[22] and parylene-C[20]. Furthermore, this can be further developed with replacement of expensive materials such as the ITO (transparent conductive electrode) alongside with the implementation of alternative transparent electrodes.

Based on the following literature and theoretical simulations analysis, mass per unit area (kg/m$^2$) and PCE are the two most significant merits of solar cell technologies regarding PPW output. Since these properties are related mainly to the device substrate, mass density, electrical, optical and physical (such as thickness) properties of active materials, transport layers and electrodes, there is still room for further development of this figure of merit (PPW). It is worth noting, that the state-of-the-art ultra-thin inorganic PVs (including CIGS and a-Si) reported PPW values of approximately 3.2 W/g (see Fig.1).[23] Nevertheless, the emerging solution-processed PV technologies, embrace the critical features of PPW applications with significant evidence to enable them to compete and outperform their inorganic counterparts. Specifically, the highest reported PPW output for the state-of-the-art emerging solution-processed solar cells is around 30 W/g and 15 W/g, corresponding to ultra-thin PSCs/OSCs and QDSCs respectively[24,25]. Thus, the emerging PVs are comparing favourably to other power-generating technologies such as electric motors (0.3 − 8.4 W/g) and heat engines/pumps (0.03 − 10 W/g) (see Fig.1),[26] and demonstrate their great potential as ultra-lightweight and efficient power sources. In this regard, we will review the literature of emerging solution-processed PVs (PSCs, OSCs and QDSCs) demonstrating PPW values, higher than their inorganic counterparts, i.e., > 3.2 W/g. Finally, in this perspective, we will explore and calculate the maximum potential of the highest-performing (OSCs and PSCs) emerging PV technologies for PPW outputs. Thus, by implementing optical and electrical simulations, we will give an accurate and deep understanding of the optimization conditions for the device's weight and PCE.

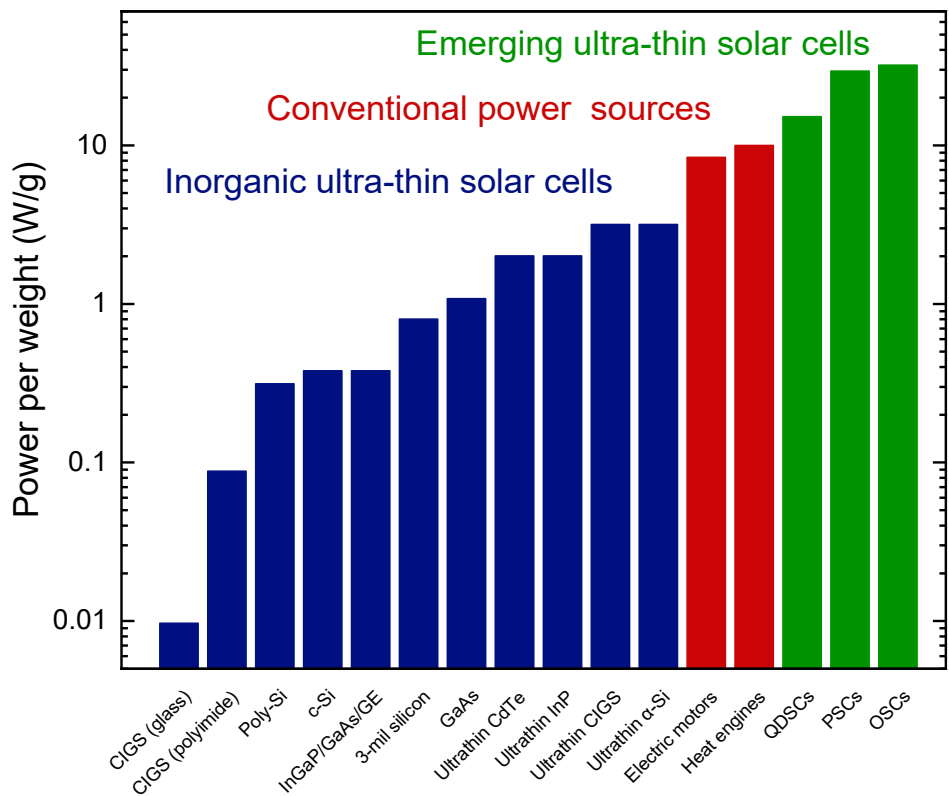

*Fig. 1.* PPW values for different ultrathin inorganic solar cells, conventional power sources and emerging ultrathin solar cells.

## 2. Literature Review of high PPW solution processed emerging PVs

The literature results of solution-processed emerging PV technologies such as PSCs, OSCs and QDSCs, has shown a great potential towards the direction of high PPW solar cells. This is because these emerging PV technologies are based on thin-film, high absorption coefficient semiconducting materials (which even with ultra-thin films they have very strong photon absorption) without the need for thick and heavy wafers as substrates. They also show promising PCEs and are compatible with low-temperature growth and deposition processes, which enables their fabrication on lightweight and ultrathin plastic substrates. On the contrary, the commercially available Silicon-based PVs, can't serve as high PPW solar cells, since their need for thick wafer-based photoactive layer (due to the indirect band-gap nature of silicon), significantly increases the total device weight, minimizing the final PPW output. Thus, ultra-thin emerging PVs, are the ideal candidates to be commercialized for high PPW solar cells applications.

## 2.1 Perovskite Solar Cells

The important properties that enable the use of PSCs for high PPW solar cells, include their extremely high efficiencies, reported even on flexible substrates(23%)[27] in combination with the relatively thin total device thickness (< 1μm, excluding the substrate thickness). Thus, thinning down the substrate thickness (reducing its weight) would enable record-high PPW solar cells.

Back in 2015, M. Kaltenbrunner et al.[26] reported the fabrication of an ultrathin PSC with a total thickness of the device on a 1.4 μm-thick PET substrate. The device configuration of PET/PEDOT:PSS/$CH_3NH_3PbI_{3-x}Cl_x$/PTCDI/$Cr_2O_3$/Cr/Au achieved a PCE value of 12% after modification of PEDOT:PSS with dimethylsulphoxide, which resulted in a pinhole-free perovskite film. Furthermore, the introduction of chromium oxide/chromium interlayer resulted in a stable device after 4 days of continuous illumination at maximum power point tracking due to the resistive nature of chromium oxide. In addition, the ultrathin PSC endured repeated compression by 25% after 100 cycles while being able to operate at a linear compression of 50%, which is equivalent to 100% tensile strain, without detriment. Finally, with a PPW value of 23 W/g, the authors demonstrated the viability of the ultrathin PSC foils for aeronautic applications with a lab-scale operational self-powered aeroplane model.

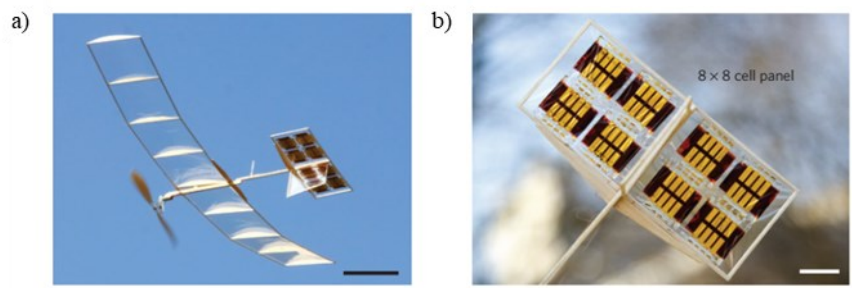

**Fig. 2.** *a) Representation of self-powered PSC model plane, b) high PPW PSCs (8x8 cells panel) integrated at the back its tail. Reproduced from ref.* [26]*with permission from Nature Publishing Group, copyright 2015.*

One year later, Z. Liu et al.,[28] demonstrated the fabrication of ultrathin flexible PSCs using chemical-vapour deposited (CVD) graphene as transparent electrodes. The structure of the flexible device (PET)/graphene/poly(3-hexylthiophene) (P3HT)/ $CH_3NH_3PbI_3$/$PC_{71}BM$/Ag was on 20μm-thick PET substrate, exhibiting PCE of 11.5%. Furthermore, the devices demonstrated bending durability, operating at different bending radii, and air stability with a PPW output of 5.07 W/g. The significantly lower PPW compared to the previous study can be explained by the much thicker PET substrate (20 μm) used in this work which significantly increased the device's weight. The ability of the devices to operate at different bending radii is attributed to the high bending durability of the transparent graphene electrodes.

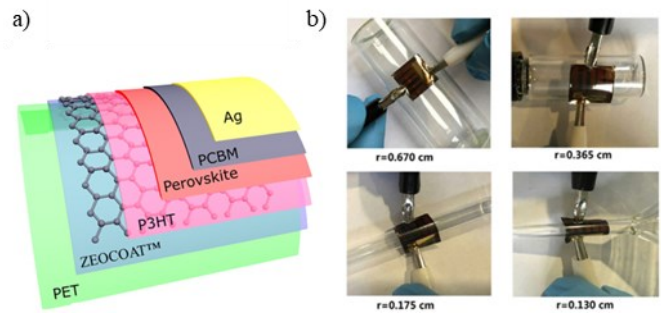

**Fig. 3.** *a) Schematic representation of the flexible PSCs, b) Bending operational tests at different radii of the reported PSCs. Reproduced from ref.* [28] *with permission from Elsevier, copyright 2016.*

Moreover, in 2017 H. Zhang et al,[29] reported the fabrication of high-performance PSCs by utilising pyridine, which promoted the formation of a high-quality perovskite film at room temperature. The champion device exhibited a PCE of 17.1% and stability up to 1000 hours of exposure to an ambient environment on a rigid substrate. Furthermore, transferring this new solution-processed room temperature method on a 1.4 μm-thick PET substrate, a flexible PSC was fabricated with the structure of PET/ITO/NiOx/CH$_3$NH$_3$PbI$_3$/C$_{60}$/Bis-C$_{60}$/Ag demonstrating PCE of 14.19% and a PPW output of 23.26 W/g. This competitive PPW value can be explained by the ultra-thin PET substrate (1.4μm) and the relatively high PCE of this ultra-thin PSC.

In 2017, Q. Wang et al.,[30] demonstrated the implementation of 4-bromobenzoic acid self-assembled monolayers (SAMs) as a passivation layer on top of NiO$_x$ for the fabrication of high-efficiency PSC at low temperatures. The SAM-treated device with the structure of PET/ITO/NiO$_x$/SAM/CH$_3$NH$_3$PbI$_3$/PCBM/BisC$_{60}$/Ag resulted in a PCE of 16.2%, mainly due to the positive dipole moment of SAM which induced a band bending downward at the interface of NiOx nanoparticles and the perovskite film, minimizing the energy level offsets. The flexibility of the device was investigated, and it was found that the devices retained 75% of their initial PCE after 110 bending cycles, proving decent mechanical stability. Finally, the

PPW value of the flexible PSCs was calculated as high as 26.9 W/g. This high PPW was mainly due to the high PCE of the PSCs.

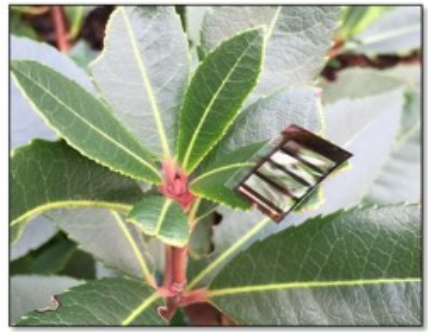

**Fig. 4.** *Photograph of the fabricated ultralight, flexible PSCs on a leaf demonstrating its lightweight feature. Reproduced from ref. [30] with permission from Wiley-VCH Verlag GmbH & Co. KGaA, copyright 2017.*

Two years later, in 2019 M. Xie et al.,[31] developed super-flexible PSCs with transparent conductive electrodes based on inject-printed silver nanowires (Ag NWs) network on a 17μm-thick PET substrate. By modifying the Ag NWs electrode with ammonia and polyetherimine modified PEDOT:PSS (m-FCE) layer, a complete device structure of PET/Ag NWs/m-FCE/PEDOT:PSS/ $CH_3NH_3PbI_{3-x}Cl_x$/$PC_{61}BM$/Al was reported. The sheet resistance ($R_{sheet}$) of Ag NWs electrode (approx. 50 Ω/sq) was stable up to 5000 consecutive bend cycles. The super-flexible PSCs exhibited an optimized PCE of 13.32% and a PPW output of 4.16 W/g. The relatively low PPW of this work can be explained due to the thick (17μm) PET substrate alongside the relatively low PCE of the flexible PSCs.

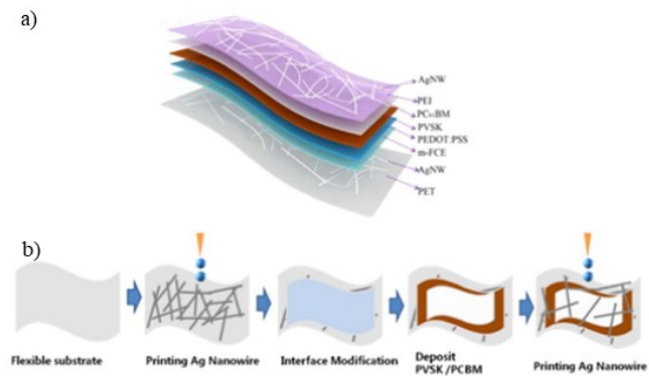

**Fig.5** *a) Device configuration of the super-flexible PSCs, b) Illustration of the fabrication of printed AgNWs bottom electrode flexible substrate. Reproduced from ref. [31] with permission from IOP Publishing Ltd., copyright 2019.*

Later this year, a highly foldable PSC was reported by H. Li et al.,[32] on a 25μm-thick cellophane substrate using a transparent oxide/ultra-thin Ag/oxide electrode with a $R_{Sheet}$ of 3.6 Ω/sq. The cellophane-based device consisting of cellophane/TiO$_2$/ultrathin Ag/TiO$_2$/C$_{60}$ pyrrolidine tris-acid (CPTA)/ CH$_3$NH$_3$PbI$_3$/Spiro-OMeTAD/Au exhibited PCE of 13.19%, value comparable to that of the reference device on glass (14.95%), which was among the highest values reported for cellophane-based PSCs. In addition, the cellophane-based devices retained over 84% of their initial PCE after 50 folding cycles with an angle of -180º and +180º, while the PCE was retained over 55% after inner and outer dual folding for 10 cycles. This high folding stability was mainly attributed to the excellent mechanical properties of cellophane substrate and the TiO$_2$/ultra-thin Ag/TiO$_2$ electrode which reduces the strain in the device when folded. Apart from folding stability, another significant advantage of cellophane substrate was the ability of shape tailoring the device with scissors and disposal via incineration. Finally, the cellophane-based PSC demonstrated a PPW output of 3.89 W/g. This low PPW is mainly attributed to the thick cellophane substrate and the gold electrode used as the metal electrode (high mass-density electrode).

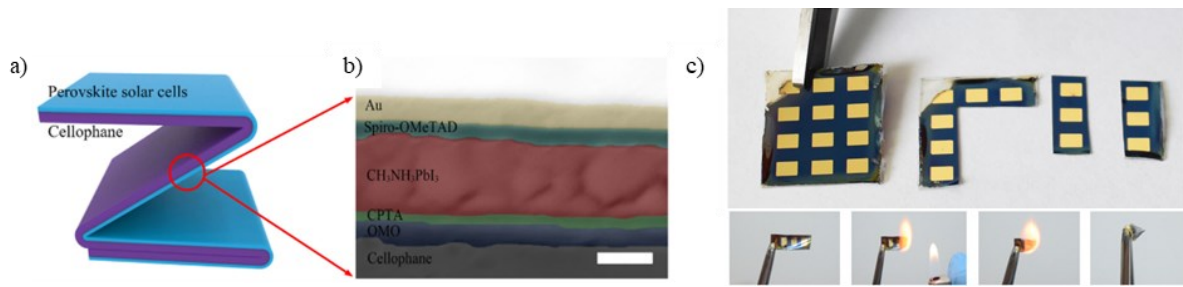

**Fig. 6.** *a) Schematic illustration of cellophane-based PSCs, b) PSCs architecture structure c) Different shapes of the fabricated PSCs demonstrated by trimming, c) Burning the PCSs on cellophane substrates. Reproduced from ref.* [32] *with permission from Wiley-VCH Verlag GmbH & Co. KGaA, copyright 2019.*

Lastly, in 2019 another study related to high PPW in PSCs was reported by S. Kang et al.[24] where ultra-light and flexible PSCs have been fabricated on 1.3μm-thick PEN foil using orthogonal Ag NWs. The device structure consisted of PEN/orthogonal AgNWs electrode/PH1000/AI4083/CH$_3$NH$_3$PbBr$_x$I$_{3-x}$(MAPbBr$_x$I$_{3-x}$)/PC$_{61}$BM/Al. The orthogonal architecture of Ag NWs resulted in a PCE of 13.05% and 12.85% for PET substrate of 120μm and PEN foil of 1.3μm, respectively. Furthermore, the devices were tested for mechanical durability retaining 80% of their initial PCE after 1000 bending cycles. The PPW output of the ultra-thin (1.3μm) and flexible perovskite solar cell was 29.4 W/g. This is the highest reported

PPW in PSCs which is achieved mainly due to the ultra-thin substrate (1.3μm), the use of aluminium electrode (low mass-density electrode) as well as the use of organic hole and electron transport layers (lower mass-density compared to the inorganic transport layers).

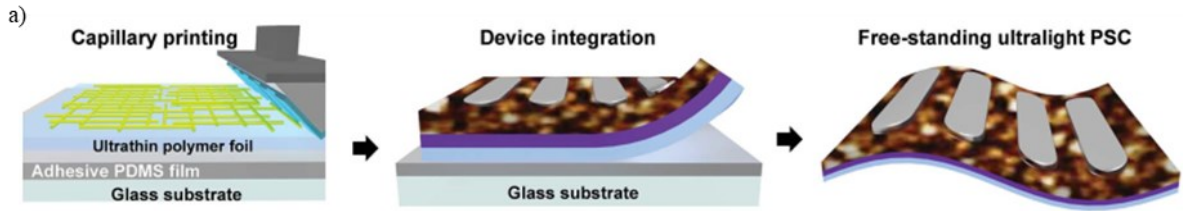

**Fig.7** *a) Representation of the fabrication procedure for the ultra-thin AgNWs electrode-based PSCs. Reproduced from ref.* [24] *with permission from Royal Society of Chemistry, copyright 2019.*

Exploring the above literature on PSCs of the reported concepts, numerous studies have demonstrated high values of PPW reaching almost 30 W/g with great operational, mechanical and strain properties, highlighting their compatibility with the area of high PPW flexible applications. One of the main challenges that need to be tackled for further improving the PPW output of PSCs, is the relatively low PCE on ultra-thin substrates (approximately 1μm thick), which are significantly lower compared to their rigid counterparts.

**Table 1.** Summary of the key parameters of high PPW PSCs reported to date.

| Device structure | Substrate | PCE (%) | PCE (%) on rigid substrate (electrode) | PPW (W/g) | Ref. |
|---|---|---|---|---|---|
| PET/PEDOT:PSS/CH3NH3PbI3−xClx/PTCDI/Cr2O3/Cr/Au | 1.4 μm PET | 12 | 12.5 (glass/ITO) | 23 | [26] |
| PET/graphene/poly(3-hexylthiophene) (P3HT)/CH3NH3PbI3/PC71BM/Ag | 20μm PET | 11.50 | 16.24 (glass/ITO) | 5.07 | [28] |
| PET/ITO/NiOx/CH3NH3PbI3/C60/Bis-C60/Ag | 1.4 μm PET | 14.19 | 17.1 (glass/ITO) | 23.26 | [29] |

| Device Architecture | Substrate | PCE (%) | Reference PCE | PPW (W/g) | Ref. |
|---|---|---|---|---|---|
| PET/ITO/NiOx/SAM/MAPbI3/PCBM/bisC60/Ag | PET | 16.20 | 18.4 (glass/ITO) | 26.9 | 30 |
| PET/Ag NWs/m-FCE/PEDOT:PSS/perovskite/PC61BM/Al | 17μm PET | 13.32 | 14.59 (glass/ITO) | 4.16 | 31 |
| cellophane/TiO2/ultrathin Ag/TiO2/C60 pyrrolidine tris-acid (CPTA)/ CH3NH3PbI3/Spiro-OMeTAD/Au | 25μm Cellophane | 13.19% | 14.95% (Glass/TiO2/ultrathin Ag/TiO2) | 3.89 | 32 |
| PEN/orthogonal Ag NW electrode/PH1000/AI4083/CH3NH3PbBrxI3-x(MAPbBrxI3-x)/PC61BM/Al | 1.3μm PEN | 12.85% | 17.51% (glass/Ag NWs) | 29.4 | 24 |

## 2.2 Organic Solar Cells

Similarly, OSCs have demonstrated the concept of high PPW as a unique feature of their class of PV technology. Due to their ultrathin and lightweight layers (organic have lower mass densities compared to inorganic materials) which results in lower device weight, the high absorption coefficients of organic semiconductors and the continuous rise of their efficiencies, OSCs could potentially report extremely high PPW values.

In the year 2012, M. Kaltenbrunner et al.,[15] demonstrated OSCs on 1.4 μm-thick PET substrates with a total device thickness below 2 μm. The prepared cells with an architecture of PET/PEDOT:PSS/P3HT:PCBM/Ca/Ag showed excellent mechanical resilience to cyclic compression–stretching with over 20 full cycles and negligible loss in device PCE. Under one-sun illumination, the device resulted in a PCE of 4.2% (fullerene-based OSCs), comparable to standard OSCs using P3HT:PCBM on ITO-coated glass substrates. The flexible polymer-based photovoltaic device demonstrated a PPW output of 10 W/g. This relatively low PPW is mainly due to the low PCE of fullerene-based OSCs at that time.

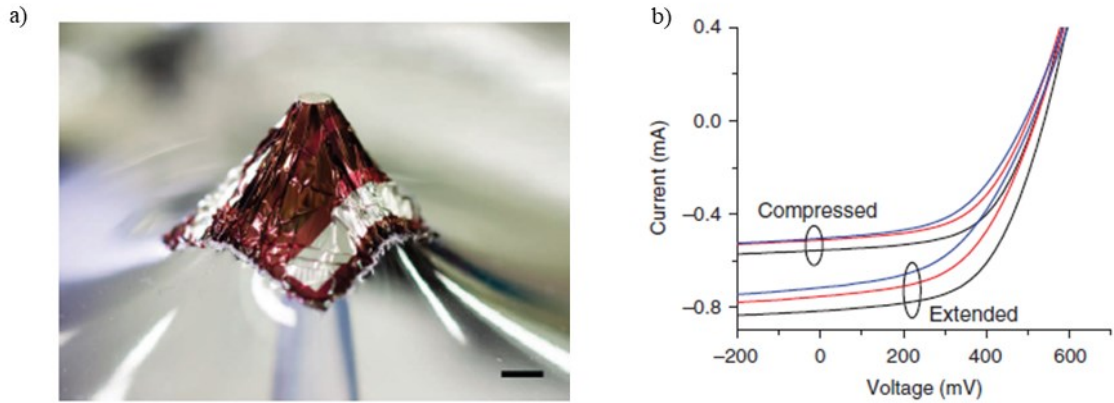

**Fig. 8.** *a) Photograph of the OSCs under extreme bending compression, b) J-V characteristic of the flexible OSCs for both compressed and extended conditions. Reproduced from ref.[15] with permission from Nature Publishing Group, copyright 2012.*

Later in 2016, J. Jean et al.,[33] followed with a study of the fabrication of an ultra-thin (1.3 μm) and lightweight (3.6 g/m$^2$) small molecule OSC, on a flexible and transparent 1 μm-thick Parylene-C substrate prepared using an in-situ vapour-phase growth. The fabricated parylene-based substrate OSC with an architecture consisting of Parylene-C/IZO/MoO$_3$/DBP/C$_{60}$/BCP/Ag showed comparable PCE to their glass-based counterparts (2.2% for the ultra-thin flexible device and 2.3% for the rigid device). The unique fabrication method ensured the avoidance of any type of contamination since the entire fabrication and encapsulation process took place under an ultra-high vacuum. Vacuum processing considers a potential candidate for handling ultra-thin flexible-based substrates OSCs with PCEs and stability similar to glass-based substrate OSCs. Using this device fabrication approach, PPW output of 6 W/g was achieved. The slightly low PPW value can be explained by the very low PCE of organic small molecule and fullerene-based OSCs. Nevertheless, with such ultra-thin substrates (1μm) the potential PPW of this architecture could be skyrocketed with an increased PCE. Thus, considering the similar thickness of alternative active materials such as non-fullerene acceptors and the recently reported novel polymer donors which output up to 19% PCE, a significantly higher PPW is expected just by replacing the suggested bilayer active layer with the recently reported novel BHJ concepts.

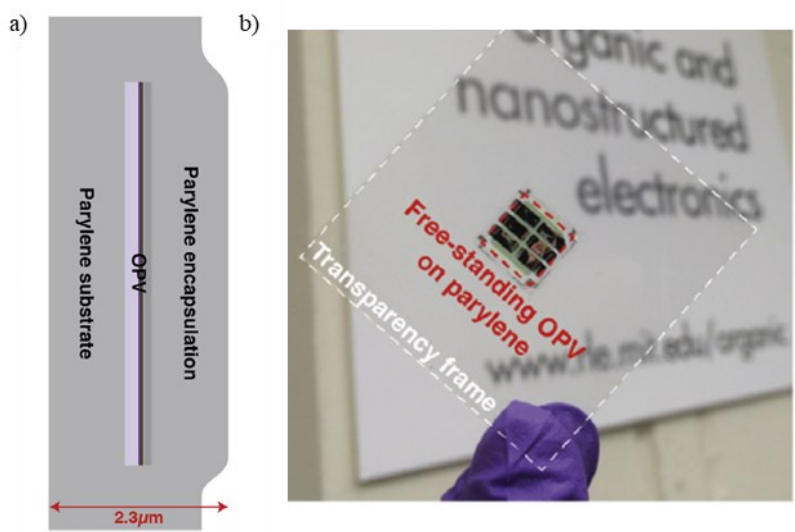

**Fig. 9.** *a) Cross section of the flexible encapsulated OSCs on parylene, b) Picture of free-standing OSC. Reproduced from ref.* [33] *with permission from Elsevier, copyright 2016.*

Highly ultraflexible OSCs and mini-modules which are mechanically robust and stable in long-term operation could ideally be integrated for future self-powered and wearable electronics. Such a concept was demonstrated in 2020 by Z. Jiang et.al,[34] with the development of an ultra-flexible 3 μm-thick non-fullerene acceptor-based OSC with a simple post-annealing process. The prepared OSCs, with the architecture of Polyimide (PI)/ITO/ZnO/PBDTTT-OFT:IEICO-4F/MoO$_X$/Ag exhibited a PCE of 13.2 % and environmental stability with 4.8% PCE loss after 4,736 hours stored in the dark at room temperature and a 20% PCE loss after 1,050 hours at 85 °C and under 30% relative humidity. The donor/non-fullerene small molecule acceptor blend in combination with the doping-induced stabilisation of the electron transport layer/ active layer interface showed an impressively long shelf lifetime with an estimated accelerated value of over 11.4 years. The small-area ultra-flexible OSC presented a 10% PCE loss after 1,000 compression/extension cycles with 37% compression. A mini-module with three series connected 0.68 cm$^2$ subcells and a total active area of 2.04 cm$^2$ reported a maximum PPW output of 9.90 W/g and mechanical stability (compression of 12%) at a minimum bending radius of less than 100 μm with a slight decrease from 11.4% to 11.0% of PCE. This is a very promising result for large areas exploitation of high PPW OSCs, considering this was the first publication reporting PPW values for solar modules (beyond the concept of the single cell OSCs). The low total device thickness (3μm) reported in this work is promising towards achieving higher PPW values by closing the PCE gap between the flexible (13.2%) and state-of-the-art rigid devices (19.2%).

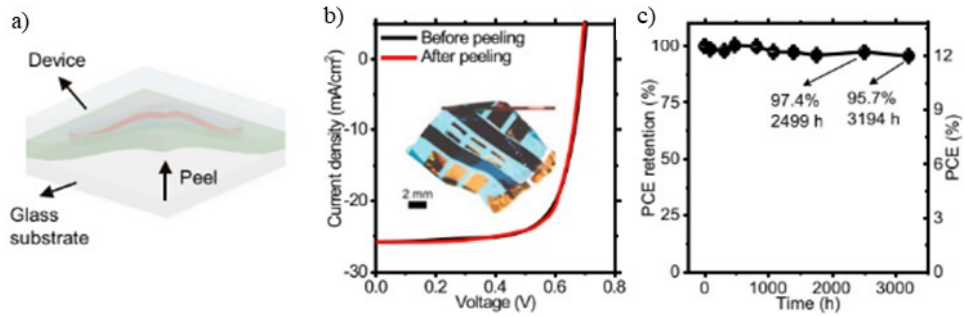

**Fig. 10.** *a) Ultra-flexible OSCs, b) J-V characteristics of the OSCs before and after the peeling process of the supporting glass substrate, c) Stability of the ultra-flexible OSCs in dark ambient conditions. Reproduced from ref. [34] with permission from National Academy of Science, copyright 2020.*

E.Bihar et al.,[35] in 2020 presented the idea of the development of an all-inkjet-printed ultrathin OSC deposited on a 1.7 μm-thick parylene substrate. The implementation of a method which enabled the cross-linking of the organic electrode PEDOT:PSS(PEDOT$_A$), preventing shunts and pinholes for the ultra-thin OSCs. Additionally, the subsequent $O_2$-plasma post-treatments on the transparent electrode ensured the control of the wettability of further layers on top of the PEDOT:PSS film. The successful deposition of all-inkjet-printed OSC with an architecture of Parylene/PEDOT:PSS(PEDOT$_A$)/ZnO/P3HT:O-IDTBR/PEDOT:PSS(PEDOT$_B$) resulted in a PCE of 3.6% and a PPW output of 6.3 W/g. After the encapsulation, the OSCs exhibited limited losses of performance after being exposed for 6 hours to different aqueous environments, in deionized water (18% loss), in phosphate-buffered saline solution (27% loss), and in seawater (22% loss). Although the thickness of the reported OSCs is promisingly low, the limited factor for the low PPW output is mainly attributed to their low PCE. Thus to further improve the PPW of this OSC architecture, alternative active layer materials should be applied to enhance their PCE. All inkjet-printing fabrication in combination with the biocompatible parylene substrate, brings OSCs one step closer to their integration into biomedical systems.

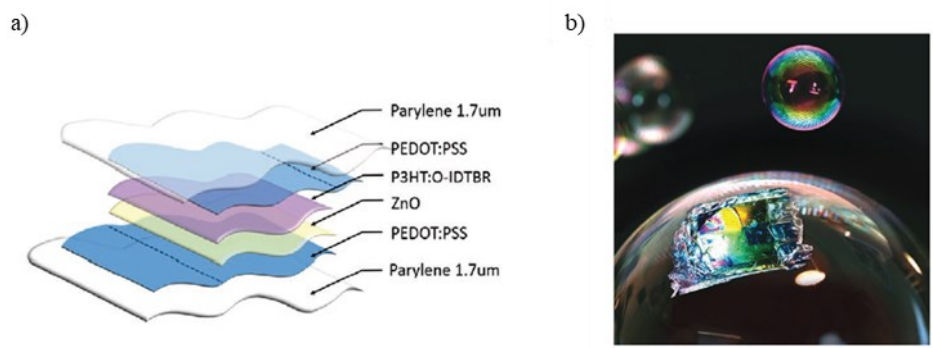

**Fig. 11.** *a) Schematic Illustration of the flexible reported OSCs, b) Photograph of the formation of ultralight inkjet-printed OSCs.embedded in a soap bubble. Reproduced from ref.* [35] *with permission from Wiley-VCH Verlag GmbH & Co. KGaA, copyright 2020.*

Another promising approach for the adaptation of the next-generation solution-processed flexible OSCs towards high PPW values is the annealing-free solution manufacturing. This concept was demonstrated in 2021, by J. Wan et al.,[36] reporting an all-annealing-free solution-processed flexible OSC deposited on a 24μm-thick, soft polyethylene (PE) substrate. The unique annealing-free fabrication process at room temperature in combination with the superacid doping and air-drying treatments of the following structure PE/PEDOT:PSS (PH1000)Anode/PEDOT:PSS/PM6:Y6/PFN-Br/Al resulted in a high PCE of 14.66%. The prepared OSCs showed a PPW output of 6.33 W/g, a light weight of 2.27 mg/cm$^2$ and an enhanced mechanical flexibility with a minimal decrease by 7.5% and 9.0% in their initial PCE after 1000 cyclic bending tests at bending radii of 0.50 and 0.25 mm, respectively. Even though the OSCs exhibited high PCE values, the reported PPW values are relatively low compared to the potential of the structure. This is explained by the thickness of the soft PE substrate (24μm thick) which increases the total weight of the device. Therefore, a straight approach to enhance the PPW is the use of a lighter and thinner substrate while keeping the PCE as high as possible.

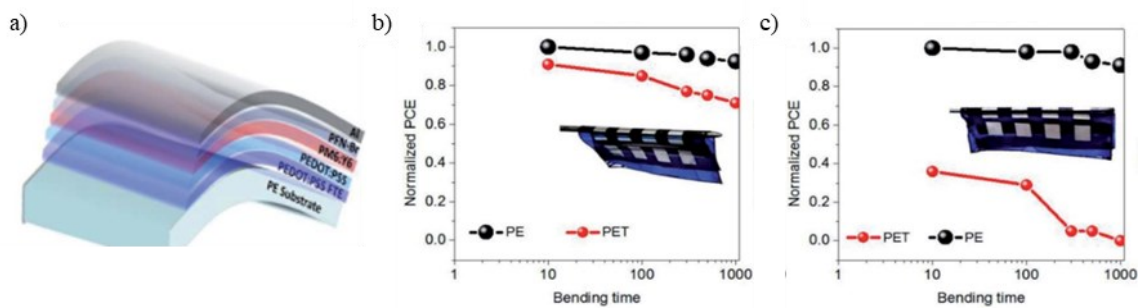

**Fig. 12.** *a) Device architecture of the PE-based substrate OSCs, b and c) normalized PCE of the flexible OSCs in bending tests at r of 0.50 and 0.25 mm. Reproduced from ref.* [36] *with permission from Royal Society of Chemistry, copyright 2021.*

Lastly, in 2021 W. Song et al.,[25] reported ultra-flexible and ultrathin binary and ternary OSCs deposited on top of 2.5 μm-thick PET substrates with high PCEs of 15.3%, 15.5% respectively. The OSCs with the architecture of PET/PEDOT:PSS(PH1000)/D18-Cl:Y6:PC$_{71}$BM/PFNDI-Br/Ag, exhibited a unique and remarkable PPW output of 32.07 W/g, which is the best ever reported for any PV technology regarding this figure of merit. Furthermore, the target device presented excellent stretch durability of 800 compression-stretching cycles with a PCE loss of

16.5% of the initial value and stored lifetime at room temperature in a nitrogen atmosphere of 1000 hours without encapsulation, retaining over 95% of its initial PCE. This ground-breaking PPW value achieved by the proposed ultra-flexible OSC is indeed remarkable, however, this could be further improved by employing a thinner substrate, and a lower mass-density metal electrode such as Al. Another way to improve the PPW of this OSC is to engineer the ultra-flexible device so that there is no or marginal PCE loss compared to the rigid device presented in this work (18%).

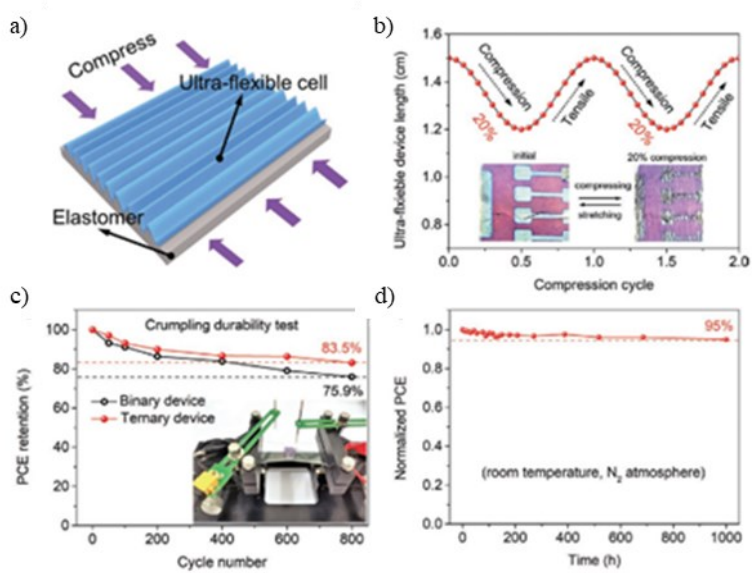

**Fig. 13.** *a) Schematic representation of the stretchable ultra-thin OSCs, b) Side length of the ultraflexible OSCs during the two compression cycles, c) Stability of the OSCs as function to the number of cycle stretching test, d) Shelf storage lifetime of the ternary OSC without encapsulation at room temperature in $N_2$ conditions. Reproduced from ref. [25] with permission from Wiley-VCH Verlag GmbH & Co. KGaA, copyright 2021.*

Ultra-flexible and lightweight OSCs embrace the major figure of merits towards the high PPW concept, due to their compatibility with low-temperature solution-processed lightweight and flexible substrates, relatively high PCE values and ultrathin transport and active materials. Exploring the above literature of OSCs, a record high PPW values have been demonstrated exceeding 32 W/g, which is the highest ever reported PPW in any PV technology. However, this PPW seems far from the optimum values, as the efficiencies of the ultrathin OSCs are significantly lower compared to their counterparts based on glass rigid substrate. Thus, a lot of attention must be devoted to engineer the flexible devices to perform comparably to the rigid devices which will further improve their PPW outputs. Additionally, alternative routes to enhance the PPW of flexible OSCs include by employing low-mass density device components (substrate, metal and transparent electrodes, active and transport layers).

**Table 2.** Summary of the key parameters of high PPW OSCs reported to date.

| Device structure | Substrate | PCE (%) | PCE (%) on rigid substrate (electrode) | PPW (W/g) | Ref. |
|---|---|---|---|---|---|
| PET (1μm)/PEDOT:PSS/P3HT:PCBM/Ca-Ag | 1.4μm PET | 4.20 | - | 10 | 15 |
| Parylene-C/IZO/MoO3/DBP/C60/BCP/Ag | 1μm Parylene-C | 2.20 | 2.30 (glass/ITO) | 6 | 33 |
| Mini module Polyimide PI/ITO/ZnO/PBDTTT-OFT:IEICO-4F/MoOX/Ag | Polyimide | 11.40 | - | 9.9 | 34 |
| Parylene/PEDOT:PSS(PEDOTA)/ZnO/P3HT:O-IDTBR/PEDOT:PSS(PEDOTB) | 1.7μm Parylene | 3.60 | 4.73 (glass/ITO) | 6.3 | 35 |
| PE/PEDOT:PSS (PH1000)/PEDOT:PSS/PM6:Y6/PFN-Br/Al | 24μm Polyethylene PE | 14.70 | 15.7 (glass/ITO) | 6.33 | 36 |
| PET/PEDOT:PSS/D18-Cl:Y6:PC71BM/PFNDI-Br/Ag | 2.5μm PET | 15.50 | 18.06 (glass/ITO) | 32.07 | 25 |

## 2.3 Quantum Dots Solar Cells

On the other hand, QDSCs have demonstrated their ability to produce relatively high PCEs, especially considering the ultrathin photoactive layer (few nanometres thick) which is very promising for applications that require lightweight devices. The first attempt to demonstrate high PPW QDSCs was reported in 2018 by X. Zhang et al.,[37] fabricating solution-processed Ag NWs electrodes on a 1.3 μm-thick PEN substrate. The flexible lightweight substrate was used for the fabrication of QDSCs with structure of PEN/Ag NWs/AZO/PbS-PbX$_2$/PbS-1,2-ethanedithol/Au, achieving a PCE of 10% and a PPW value of 15.2 W/g. Moreover, the QDSCs showed great mechanical stability maintaining their operational properties under 60% compression while maintaining 85% of their initial PCE after 40 repeated compression-stretching cycles. Finally, regarding the operational stability, retaining 85% of its initial PCE after 35 days of unencapsulated storage under ambient conditions for 290 min under 100 mW/cm$^2$ illumination at MPPT point. This relatively high PPW value is mainly attributed to their very thin active layer and substrate, however, this could significantly be improved by increasing their PCE and replacing Au electrodes with low-mass density counterparts such as Al.

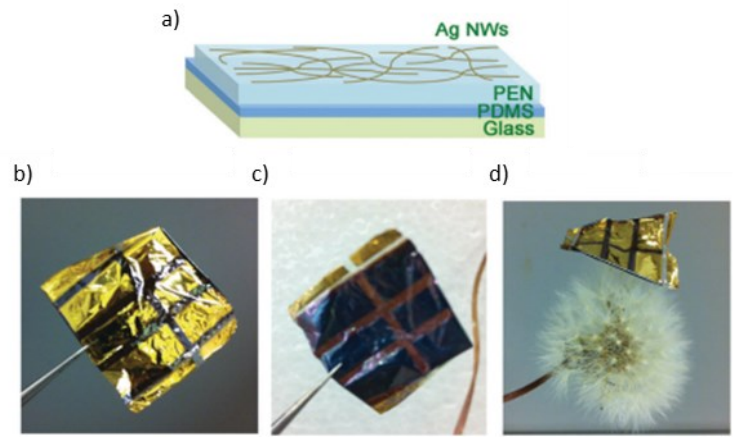

**Fig. 14.** *Representation of the transparent and AgNWs-based electrode of the device, b) and c) back and front side of the flexible QDSC, d) The flexible QDSC placed on a dandelion without any deformation of the dandelion. Reproduced from ref.* [37] *with permission from Royal Society of Chemistry, copyright 2018.*

Lately, in 2020, M.Tavakoli et al.,[38] reported the fabrication of flexible ultra-lightweight QDSCs by utilizing oxidative CVD PEDOT (oCVD) on top of CVD grown graphene (CVD-Graphene) as an interlayer to overcome the wettability issue of graphene electrodes. The 1 μm-thick parylene, combined with graphene electrodes on PET was a suitable substrate for the fabrication of flexible QDSC with a structure of PET/parylene/graphene/oCVD PEDOT/PbS QDs/LiF/Al, exhibiting a PCE value of 7.1%. After the removal of PET film, the QDSC demonstrated a PPW value of 12.3 W/g, alongside its mechanical stability, exceeding 83% of the initial PCE after 100 consecutive bending cycles. The PCE values of QDSC is the main reason for the low PPW value of the reported flexible device since the use of 1 μm-thick parylene as substrate, graphene as transparent electrode and aluminium as the metal electrode represents almost ideal cases towards lightweight device fabrication.

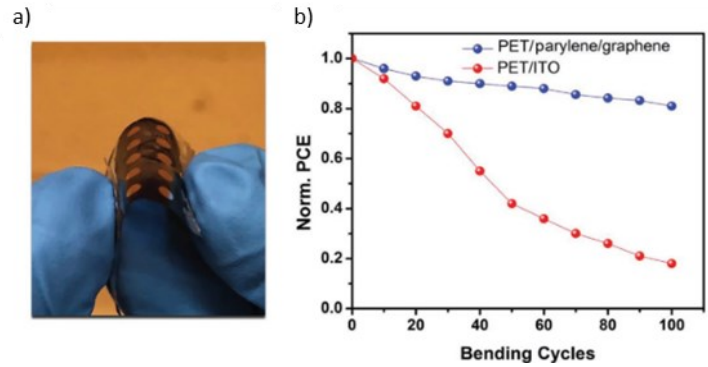

**Fig. 15.** *a) Photograph of the flexible QDSC on PET/parylene/graphene substrate, b) Normalized PCE versus bending cycles for the reference and graphene-based QDSCs. Reproduced from ref.* [38] *with permission from Wiley-VCH Verlag GmbH & Co. KGaA, copyright 2020.*

The QDSC technology could potentially be considered as an alternative candidate for high PPW applications, by addressing important limiting factors which stand along. Although they combine the essential features for lightweight and flexible concepts, their relatively low PCEs is the main parameter that needs to be optimised. Regarding substrate, electrode and transport/active materials, it has been proven that it is possible one to implement ultra-thin and lightweight materials. However, to achieve high PPW values the PCE of such lightweight QDSCs needs to be significantly improved approaching the rigid substrate certified PCE values (18.1%).

**Table 3.** Summary of the key parameters of high PPW QDSCs reported to date.

| Device structure | Substrate | PCE (%) | PCE (%) on rigid substrate (electrode) | PPW (W/g) | Ref. |
|---|---|---|---|---|---|
| PEN/Ag NWs/AZO/PbS-PbX2/PbS-1,2-ethanedithol/Au | 1.3µm PEN | 10 | 9.7 (glass/Ag NWs) | 15.2 | 37 |
| PET/parylene/graphene/oCVD PEDOT/PbS QDs/LiF/Al | 1µm Parylene | 7.10 | - | 12.3 | 38 |

# 3. Theoretical Investigation of High PPW solution-processed PVs: Requirements and Guidelines.

Based on the above-mentioned literature review, it is straightforward to conclude that among the emerging solution processed PV technologies, OSCs and PSCs present the highest potential for their implementation as power generating sources on lightweight high PPW solar applications. On the one hand, OSCs' ultrathin active layer (~100nm) and on the other hand, PSCs' high PCEs (~16%) on flexible substrates, are the key significant advantages towards the fabrication of ultra-lightweight solar cells with record PPW outputs (both OSCs and PSCs have ~30 W/g PPW output). QDSCs have a major bottleneck towards high PPW, which is the low PCEs attained (10% at maximum) that prohibits this PV technology to surpass 15 W/g PPW output. Thus, we will focus on our theoretical investigations and optimizations in both OSCs and PSCs which have the highest potential towards high PPW output.

As it has been already mentioned earlier in this article, high PPW require minimal weight, high PCE, and mechanical resilience. The relative impact of ultrathin PSCs and OSCs PCE and weight on PPW (as well as their interplay) have yet not been examined and optimized theoretically to understand their maximum potential. Therefore, OSCs and PSCs optimal design parameters, i.e., optimal layer thickness and material, for maximum PPW remain unclear. To answer these questions in the present section hence providing guidelines for designing solar cells with high PPW beyond the state-of-the-art presented in the literature, we employed an analytical, generic optical-electrical model. Using this model, we calculate the OSCs and PSCs PPW as a function of the devices' layers thickness and materials and perform an optimization procedure.

Optically, we calculated the solar cells' absorptivity by employing the transfer matrix method and used it as an input in an analytical electrical model based on current−voltage characteristics to evaluate the output power. Specifically, we extended the theoretical Shockley-Queisser model[39] to include nonradiative recombination, shunt resistance, as well as that not all the incident solar light above the bandgap leads to photocurrent (optical losses, External Quantum Efficiency (EQE) < 1). Moreover, we included different recombination mechanisms for different solar cells (OSCs and PSCs). To model the OSCs, we take radiative bimolecular recombination ($J_{bi}$) into account [40] while for the PSCs, we assume dominating Shockley-Read-Hall (SRH) nonradiative recombination by the space charge limited region ($J_{SRH}$), since most of the perovskite layer is depleted [41] the current−voltage characteristics are then given by the following diode equation[42]

$$J(V) = J_{SC} - J_{bi,0}(V)\left(e^{(qV/k_B T)} - 1\right) - J_{SRH,0}(V)\left(e^{(qV/2k_B T)} - 1\right) - \frac{V}{R_{sh}}, \quad (1)$$

where $J$ is the output current density generated by the solar cell [in A/m$^2$], $J_{SC}$ is the current density flowing at short-circuit conditions under the sun illumination, $q$ is the elementary charge of an electron, $k_B$ is Boltzmann's constant, $V$ is the output voltage, and $T$ is the temperature of the solar cell. The second and the third terms correspond to the radiative bimolecular and nonradiative SRH recombination current densities, respectively, with the corresponding dark-saturation current densities $J_{bi,0}$ and $J_{SRH,0}$ and ideality factors of 1 and 2, respectively, and the last term is due to shunt resistance (see Ref.[42] or a full description of the model). We calculate then the PCE at the maximum power point (mpp) [42] of the solar cell by PCE($V_{mp}$) = $J(V)V|_{mp}/\int I(\lambda)d\lambda$, where $I(\lambda)$ is the solar illumination represented by the measured sun's radiation, the AM 1.5G spectrum (1000 W/m$^2$), and $\lambda$ the wavelength of light. PPW is then given by PPW = $J(V)V|_{mp}/w_g$, where $w_g$ is solar cell's weight (in g) calculated by obtaining the mass density (in g/m$^2$) of each material of the ultrathin solar cell from Ref.[43] the thickness of each layer, and assuming an area of 1 m$^2$.

By comparing each layer's thickness and material requirements for maximum PCE versus maximum PPW, one can evaluate the relative impact of solar cells' PCE and weight on PPW output, hence determining optimum design parameters for maximizing PPW output. Therefore, the optimization of the thin-film stack is implemented by combining a global evolutionary algorithm ("Genetic" algorithm) and a local optimization method ("Nelder Mead Simplex") [42,44] over two objective functions, PCE and PPW output. The independent variables (optimized according to the objective function) are the thicknesses of the active, hole- (HTL), and electron-transporting (ETL) layers, while the substrate, transparent-, and metal-electrode thicknesses are fixed to preserve low sheet resistance and high stability, and mechanical protection.

## 3.1 Theoretical simulation and optimization of PSCs' PPW

We start our analysis by examining the PSCs with the highest reported PPW of an inverted stack layout (upside-down fabrication on top of an ultra-thin, light, and flexible plastic substrate).[24] The detailed structure of the device is as follows (see also Fig.16 a), polyethylene naphthalate (PEN – 1.3 μm)/silver nanowire network (AgNWs/PH1000 − 70 nm)/PEDOT:PSS (30–60 nm)/perovskite (CH$_3$NH$_3$PbI$_3$ – 350–600 nm)/PCBM (30–50 nm)/Al (100 nm), where the numbers indicate the thickness of each layer, and the ranges indicate the boundary values of the independent variables corresponding to typical thicknesses in experiments. To model the AgNWs network, we assume a thin polymer binder (70 nm) containing nanowires so that the weight fraction of nanowire to polymer is 30%, with material data from Ref.[45] The simulated $J$–$V$ curves corresponding to the optimized PSCs for maximum PCE (black) and maximum PPW (red) and their PV characteristics, i.e., $J_{SC}$, open-circuit voltage, $V_{OC}$, fill factor, PCE, PPW, and $w_g$, are shown in Fig. 16 e.

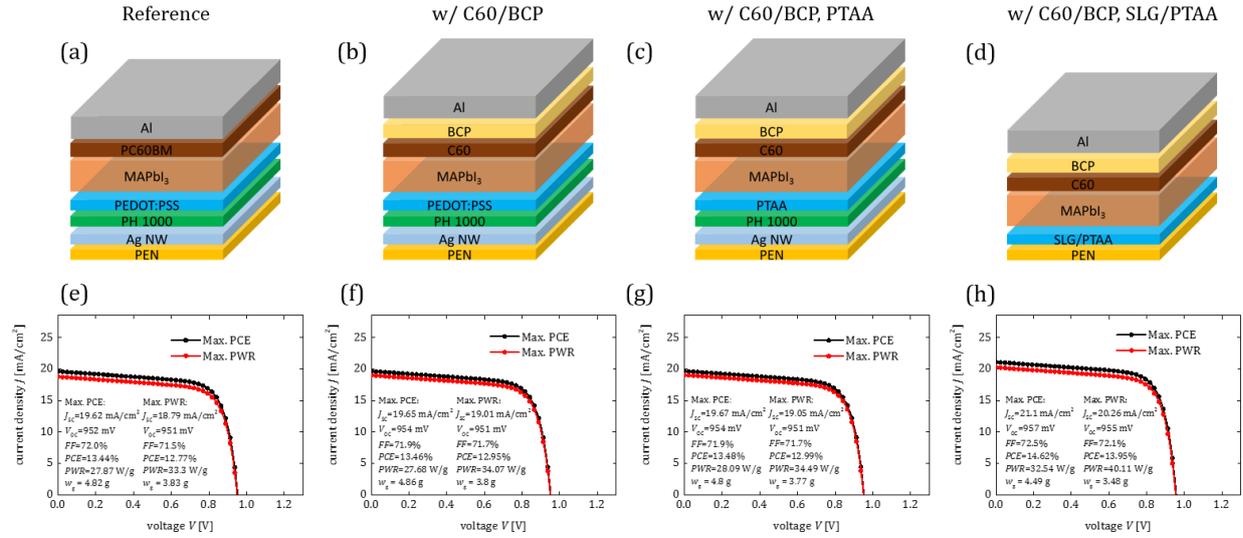

**Fig. 16.** *Geometry (a–d) and J–V characteristics (e–h) of the reference (a, e) and proposed (b–d, f–h) ultra-lightweight PSCs, and the corresponding PV characteristics ($J_{SC}$, $V_{OC}$, FF, PCE, PPW, and $w_g$). In (e–h), the black lines correspond to the optimized PSCs for maximum PCE and the red lines to the optimized PSCs for maximum PPW.*

First, the simulated PPW of the optimized PSC for maximum PCE and PPW equals ~27.9 and ~33.3 W/g, respectively, very close to the one calculated experimentally (~29.4 W/g).[24] Notably, the optimum layer thicknesses for high PPW differ from those for high PCE. For optimal PCE, the optimized perovskite layer thickness almost matches the highest boundary value of 600 nm (590 nm), while for optimal PPW, the optimised perovskite layer thickness equals the minimum boundary value of 350 nm (350.1 nm) despite the decreased $J_{SC}$, hence PCE. Moreover, for optimal PPW output, the optimized thicknesses of the ETL and HTL also almost match the minimum boundary values of 30 nm (i.e., 30.4 and 37.4 nm, respectively), while for optimal PCE, they take values dictated by interference effects in the thin-film stack (i.e., 35.4 and 49.9 nm, respectively).

These results indicate that for maximum PPW output, the layers of the PSC should be as thin as possible to decrease its weight (up to a limit not to affect devices PCE due to poor absorption and charge extraction). In contrast, for maximum PCE, the perovskite layer should be as thick as possible (up to a certain limit as shown above, not to affect the photoexcited charge carriers extraction) to maximize the photocurrent, hence its PCE. Specifically, the optimized PSC for maximum PCE has a ~0.67 % higher PCE than the optimized PSC for maximum PPW (i.e., PCE = 13.44 versus 12.77 %), but a ~5.43 W/g lower PPW (i.e., PPW = 27.87 versus 33.3 W/g) due to the excessive weight ($w_g$ = 4.82 versus 3.83 g). Consequently, the impact of solar cell weight decrease with decreasing layers' thickness (~1 g) is more important to PPW than the relative PCE improvements with increasing perovskite's layer thickness (~5.25 %).

To further increase the PPW (beyond the state-of-the-art), we examined decreasing PSCs' weight by integrating more efficient, stable, thinner, and lighter materials. First, we examined the impact of substituting the flexible transparent AgNWs electrode, PEDOT:PSS hole-, and $PC_{60}BM$ electron-transporting layer on PPW. We utilize PTAA and $C_{60}$/BCP of lower typical thicknesses providing improved PCE and stability over PEDOT:PSS, and $PC_{60}BM$ counterparts. Specifically, the thickness boundary values for PTAA, $C_{60}$, and BCP are 5–15 nm, 10–30 nm, and 5–7 nm, respectively. Consequently, integrating PTAA and $C_{60}$/BCP into the solar cell, the PSC weight decreased from 3.83 to 3.77 g, and $J_{SC}$ slightly increased, leading to a PPW increase of 1.2 W/g (from 33.3 to 34.39 W/g), see Fig. 16c and g. The main PPW improvement resulted from the substitution of the AgNWs electrode. Till date, several approaches for the fabrication of high-performance flexible PSCs using various alternatives to conventional indium tin oxide (ITO) electrode, of a more brittle nature, such as metal nanowires (AgNWs), carbon nanotubes (CNTs), graphene, and metal mesh, have been reported.[24] Among these, AgNWs network have been used as an attractive flexible transparent electrode owing to their excellent electrical and optical properties, mechanical flexibility, and fabrication simplicity, e.g., through conventional solution-based processes. However, low transmittance and high surface roughness, related to AgNWs aggregation, limit their application in high-performance flexible PSCs.[24] In our study, we substitute the ~70-nm-thick AgNWs network/PH1000 with a single-layer graphene (SLG) to enhance the transmittance in visible and decrease the PSC weight simultaneously. SLG is naturally flexible, ultrathin (~0.5 nm), and with high mechanical robustness, especially compared to conventional ITO and is an excellent match with PTAA. Consequently, integrating SLG in the PSC, its weight decreased substantially (from 3.77 to 3.48 g), and $J_{SC}$ increased (from 19.05 to 20.26 mA/cm$^2$), leading to a PPW increase of 5.72 W/g (from 34.49 to 40.11 W/g), showing a high-PPW potential for PSCs with graphene-based electrodes and a high room for improvement.

In all cases (see Fig. 16), decreasing PSCs' weight with decreasing layers' thickness is more significant to PPW than the relative PCE improvements with increasing perovskite thickness. These results indicate that for maximum PPW for PSCs, the layers of the PSC should be as thin as possible to decrease its weight. In contrast, for maximum PCE, the perovskite layer should be as thick as possible (up to 600 nm) to maximize the photocurrent.

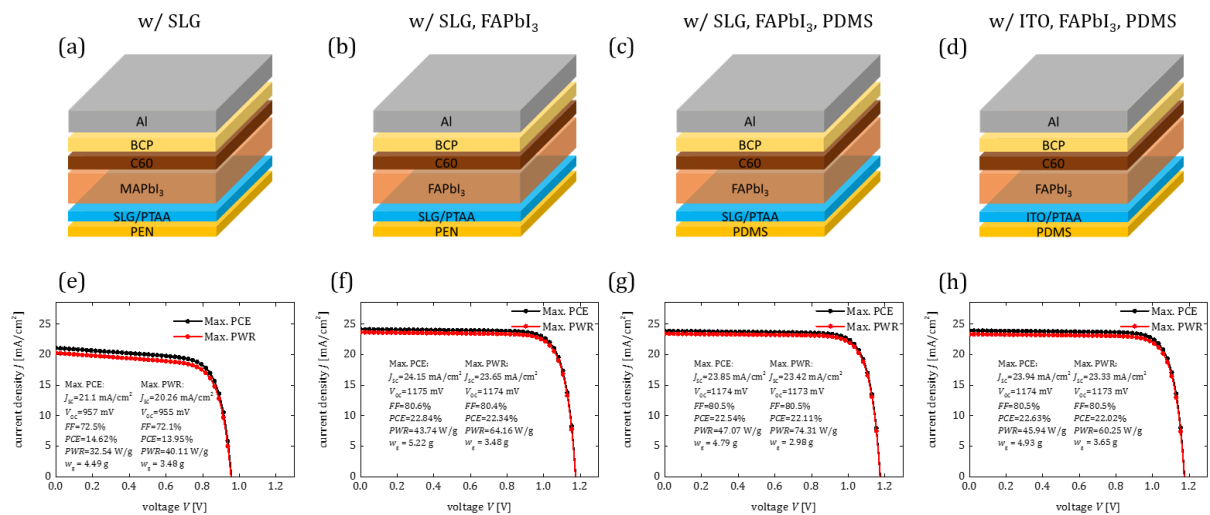

**Fig. 17.** *Geometry (a–d) and J–V characteristics (e–h) of the proposed (a–c, e–g) and with conventional ITO (d, h) ultra-lightweight PSCs, and the corresponding PV characteristics ($J_{SC}$, $V_{OC}$, FF, PCE, PPW, and $w_g$). In (e–h), the black lines correspond to the optimized PSCs for maximum PCE and the red lines to the optimized PSCs for maximum PPW.*

Besides PSCs' weight decrease, increasing PSCs' PCE beyond the state-of-the-art is expected to further increase PPW. Recently, there was a significant progress in enhancing PSCs' PCE.[46] Until recently, $CH_3NH_3PbI_3$ perovskite was extensively examined for rigid and light-weight solar cells due to its fabrication simplicity, while its direct band-gap of ~1.55 eV (~800 nm, i.e., at the onset of the optical range), very close to the ideal compared to other perovskites in which another halide is present, led to high PCEs. Recently, formamidinium lead iodide ($FAPbI_3$) has endowed PCEs up to 26% because of its optimal bandgap, enhanced transport properties, and thermal stability.[46–48] Indicatively, studies have utilized $FAPbI_3$-based PSCs with active layer thicknesses as thick as 800 nm without decreasing the cells' quantum efficiency, greatly enhancing PSCs' $J_{SC}$ and $V_{OC}$, hence PCE.[46] Therefore, in our study, we used $FAPbI_3$ as the absorber layer, assuming a higher boundary value of $FAPbI_3$ thickness than 600 nm of 350–800 nm.

Interestingly, the use of FAPbI$_3$ led to a high PCE improvement of ~60% (from PCE = 13.95 to 22.34%) due to improved $J_{SC}$, $V_{OC}$ and FF (see Fig. 17c,e and f). As a result, PPW increased substantially by ~24.1 W/g (from PPW = 40.1 to 64.2 W/g) (see insets in Fig. 17e and f). Moreover, in agreement with the results of Fig. 16 for CH$_3$NH$_3$PbI$_3$, the results of Fig. 17f indicate that the optimal FAPbI$_3$ thickness for the optimized PSC for maximum PCE is almost maximum (~781 nm). While the optimal FAPbI$_3$ thickness for the optimized PSC for maximum PPW almost matches the minimum boundary value (~351 nm).

Ultra-thin, light, and flexible plastic substrates also play a critical role in PPW output. They comprise the thicker part of an ultra-lightweight solar cell (with typical thicknesses of ~1.3 μm) to provide sufficient mechanical support and protection against adverse environmental influences such as oxygen and water vapor transmission. Therefore, to further increase PPW, we substituted the poly(ethylene naphthalate) (PEN) substrate with a common inexpensive polydimethylsiloxane (PDMS) ultra-thin substrate (~1.3 μm) of lower mass density (i.e., 0.97 compared to 1.36 g/cm$^3$ of PEN.[43] As expected, PPW increased substantially by ~10.1 W/g (from 64.2 to 74.3 W/g), leading to a theoretically calculated and realistic PPW of 74.3 W/g (with materials utilized in literature), considered an upper theoretical PPW limit for PSCs.

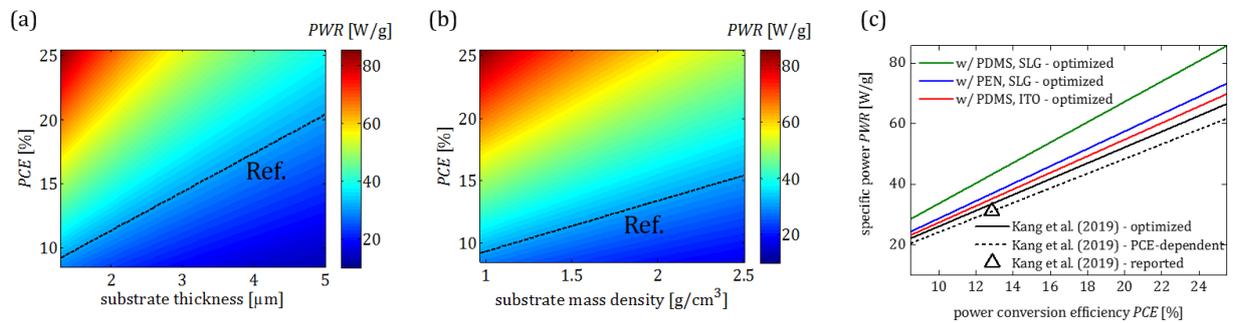

**Fig. 18.** *PPW as a function of (a) substrate thickness (1.3–5 μm), (b) 1.3-μm-thick substrate mass density (0.97–2.5 g/cm$^2$), and PCE (up to 25.5%, i.e., close to the highest reported for PSCs) for the optimized PSC shown in Fig. 17c. The black line corresponds to the highest reported PPW for PSCs plotted for reference.[24] (c) PPW as a function of PCE for different PSCs configurations (corresponding to different solar cell weight – see Fig. 17). The black triangle corresponds to the PSC with the highest reported PPW[24], while the dashed black line indicates its PCE-dependent PPW plotted for reference.*

In Fig.18a and b, we show the impact of the substrate on the PPW of PSCs. Specifically, in Fig.18a, we plot PPW as a function of PCE (up to 25.5%) and substrate thickness (in the range of 1.3–5 μm), assuming the optimized PSC with PDMS (see Fig. 17c). In Fig.18b, we plot

PPW as a function of PCE and substrate mass density (from 0.97 to 2.5 g/cm$^2$, corresponding to PDMS and glass, respectively), assuming a typical ultra-thin 1.3-μm-thick substrate. The black line corresponds to the highest reported PPW for PSCs plotted for reference. As seen in Fig. 18a and b, for an optimal PSC composed of materials shown in Fig.17c, even a PCE of ~10% is enough to surpass the record PPW reported in literature to date. Interestingly, even increasing the substrate thickness or assuming a superstrate encapsulant, i.e., for enhancing solar cell stability and mechanical support, and different substrate material (see Fig.18b), there is still significant room for improvement.

In Fig.18c, we plot the PPW as a function of PCE for different PSCs configurations (corresponding to different solar cell weight – see Fig. 17) to show the material and thickness optimization impact on PPW. The black triangle corresponds to the PSC with the highest reported PPW output in the literature[24], while the dashed black line indicates its PCE-dependent PPW plotted for reference. The solid black line indicates the PCE-dependent PPW with optimized layer thickness values (see Fig. 16a and e). We also plot for reference the PCE-dependent PPW of an optimized PSC with a conventional ITO electrode (red line), see Fig. 17d and h. Even assuming the same PCE, integrating into the solar cell lighter and efficient materials such as ultrathin electrodes (i.e., single-layer graphene) and substrates of lower mass density (i.e., PDMS), a high increase in PPW is expected compared to the state-of-the-art (~10 W/g for PCE = 12.85% – see green versus black line), which further increases as PCE increases (~20 W/g for PCE = 25.5%). Interestingly, by substituting the conventional ITO with an ultrathin electrode (SLG), an 8–16 W/g increase in PPW is expected (see green versus red line).

## 3.2 Theoretical simulations and optimization of OSCs' PPW

Besides PSCs, OSCs have emerged as highly promising, ultra-lightweight, flexible, and highly efficient PV technology. This is mainly due to their lower mass density (i.e., active layer of ~1.3 compared to ~4 g/cm$^3$ for $CH_3NH_3PbI_3$),[25,43] low-temperature processing compatibility, i.e., facilitating the employment of ultrathin, light, and flexible plastic substrates such as PET, PEN foils, Parylene C, and PI, and recent improvements in their PCE (due to improved exciton disassociation and charge transport), leading to high related PPW values.[25]

Specifically, the record-reported PPW for an OSC of conventional device stack (upside-down fabrication on top of an ultrathin, light, and flexible PET substrate) equals ~32.1 W/g at a weight of 4.83 g/m$^2$ and a PCE of 15.5% (~2.7 W/g higher than the record PPW of PSCs).[25] The detailed structure of the device is as follows: PET – 2.5 μm/PH1000 (40 nm)/PEDOT:PSS

(30 nm)/active layer (D18-Cl:Y6:PC$_{71}$BM – 100–300 nm)/ PFNDI-Br (5–30 nm)/Ag (150 nm).[25]

In our theoretical study, we proposed and simulated a similar ternary OSC as an ultra-lightweight solar cell based on PM6:D18:L8BO instead of D18-Cl:Y6:PC$_{71}$BM active layer, that recently led to world record certified PCE of 19.2%.[40] The detailed structure of the proposed device is as follows (see also Fig. 19a): PDMS (1.3 μm)/PH1000 (40–80 nm)/PEDOT:PSS (20–40 nm)/active layer (PM6:D18:L8BO – 90–180 nm)/ PNDIT-F3N (5–10 nm)/Aluminum (Al – 150 nm), where the numbers indicate the thickness of each layer and the ranges indicate the boundary values of the independent variables corresponding to typical thicknesses in experiments. Note that the Ag electrode utilized in Ref.[40] is substituted by an Al electrode of much lower mass density (i.e., ~2.7 compared to ~10.49 g/cm$^3$ of Ag [43] and the 2.5-μm-thick PET substrate by a lighter, thinner 1.3-μm-thick PDMS substrate of lower mass density (i.e., ~0.97 compared to ~1.39 g/cm$^3$ of PET [43] to enhance PPW. The related $J-V$ curves corresponding to the optimized OSCs for maximum PCE (black) and maximum PPW (red) and their PV characteristics, i.e., $J_{SC}$, $V_{OC}$, FF, PCE, PPW, and $w_g$, are shown in Fig. 19c.

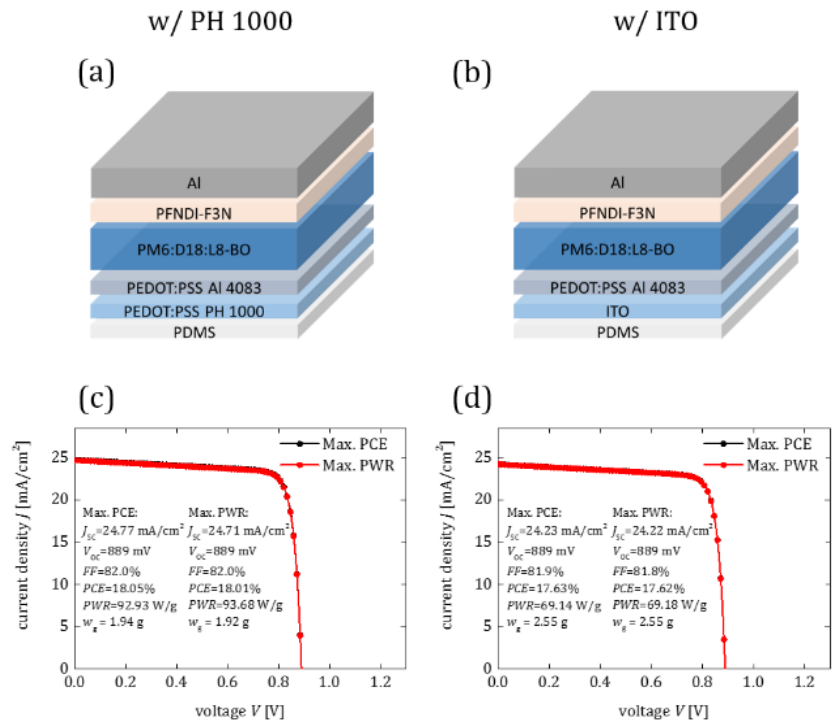

**Fig. 19.** *Geometry (a−b) and J−V characteristics (c–d) of the proposed (a, c) and with conventional ITO (b, d) ultra-lightweight OSCs, and the corresponding PV characteristics ($J_{SC}$, $V_{OC}$, FF, PCE, PPW, and $w_g$). In (c–d), the black lines correspond to the optimized OSCs for maximum PCE and the red lines to the optimized OSCs for maximum PPW.*

Surprisingly, due to ultra-lightweight OSC's high PCE (~18%) and active layers' low mass density, PPW output reached extremely high values, up to ~93.7 W/g (see inset in Fig. 19c), indicating a very high PPW potential for OSCs and room for improvement. Specifically, realistic optimized OSC's weight equals ~1.92 g (Fig. 19c), compared to 2.98 g of the optimized PSC (Fig. 17g). As a result, and despite OSCs' lower PCE compared to PSCs, they exhibited much higher PPW and therefore seem more promising for PV applications with high PPW output. Moreover, optimized (for maximum PCE and PPW) OSCs' $J-V$ curves converge (see black versus red curves in Fig. 19c) in contrast to PSCs (Fig. 16), indicating similar thickness requirements for optimal PCE and PPW simultaneously. The results in Fig. 19c indicate that PPW may increase substantially by ~61.6 W/g (from PPW = 32.1 to 93.7 W/g), leading to a maximum value of 93.7 W/g, considered an upper theoretical PPW limit for OSCs with materials utilized in the literature.

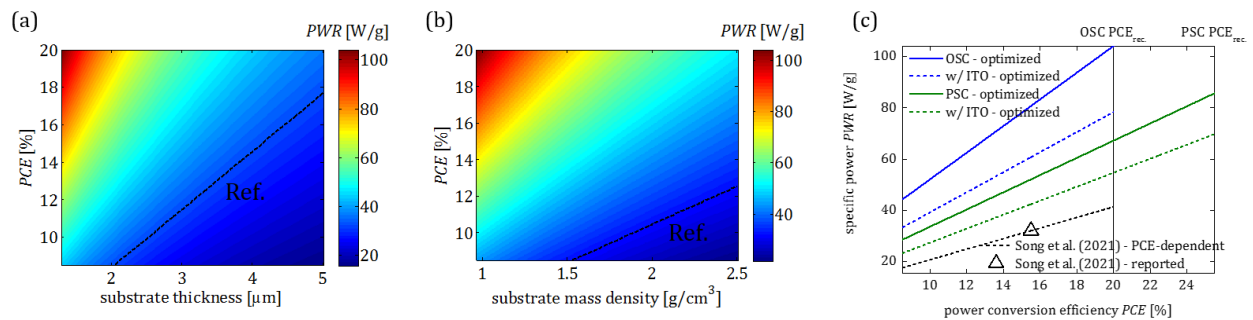

**Fig. 20.** *PPW as a function of (a) substrate thickness (1.3–5 μm), (b) 1.3-μm-thick substrate mass density (0.97–2.5 g/cm$^2$), and PCE (up to 20%, i.e., the highest reported for OSCs) for the optimized OSC shown in Fig.19a . The black line corresponds to the highest reported PPW for OSCs plotted for reference. (c) PPW as a function of PCE for different optimized OSCs (see Fig. 19) and PSCs configurations (see Fig. 17c and d). The black triangle corresponds to the highest reported PPW for solar cells, corresponding to ultra-lightweight ternary OSCs [25] while the dashed black line indicates its PCE-dependent PPW plotted for reference.*

The impact of the substrate thickness and mass density on the PPW output of OSCs is shown in Fig. 20a and b, respectively. The black line corresponds to the highest reported PPW for OSCs (~32.1 W/g) plotted for reference. As seen in Fig. 20a and b, for an OSC with materials as shown in Fig. 19a, even a PCE of ~6.2% is enough to surpass the record PPW for ultra-light weight OSCs. Interestingly, even increasing the substrate thickness or assuming a superstrate encapsulant, i.e., for enhancing solar cell stability and mechanical support, and different substrate material (see Fig. 20b), there is still significant room for improvement, even more than PSCs, despite their higher PCE (see also Fig. 18a and b).

This is better illustrated in Fig.20c, where we plot the PPW as a function of PCE for different optimized OSCs (see Fig.19) and PSCs configurations (see Fig.17c and d) and show the impact of solar cells' material and thickness optimization on PPW. The black triangle corresponds to the highest reported PPW for solar cells, corresponding to ultra-lightweight OSCs [25] plotted for reference. Despite OSCs lower PCEs compared to PSCs (i.e., up to ~20% for OSCs versus ~25.5% for PSCs [40,46] their PPW output is higher (by ~8.7 up to ~18.5 W/g) due to their lower mass density. Additionally, the impact of substituting conventional ITO in OSCs on PPW output is higher than in PSCs (see blue versus green lines), despite optimized OSC's thicker PH1000 electrode (~40 nm – see Fig. 4a) than optimized PSC's single-layer graphene electrode (~0.5 nm – see Fig. 17c). Specifically, by substituting the conventional ITO with a thin PH1000 electrode (~40 nm), an 11–25 W/g increase in PPW is expected (see blue dashed versus blue solid line). These results indicate that OSCs seem more promising than PSCs for PV applications with high PPW output.

Summarizing, the results in the present section indicate an upper theoretical PPW limit of 74.3 and 93.7 W/g for PSCs and OSCs, respectively, with materials utilized in literature, indicating significant room for improvement. Additionally, OSCs seem more promising than their counterpart PSCs for PV applications with high PPW output due to their low mass density. Moreover, our results indicate, that for maximum PPW with PSCs, their layers should be as thin as possible (keeping at the same time the PCE as high as possible) to decrease the solar cell weight. In contrast, we found similar thickness requirements for optimal PCE and PPW for OSCs owing to the active layer's low mass density. Finally, emerging, and highly efficient active layer materials, such as $FAPbI_3$ and PM6:D18:L8BO, lighter substrates, such as PDMS, thinner transparent electrodes, such as SLG and lighter metal electrodes, such as Al seem promising for substantially increasing solar cells' PPW beyond the state-of-the-art.

## 4. Perspective and conclusions

Before summarizing and concluding this perspective article, it is important to point out the competitive advantages and disadvantages of both solution-processed PV technologies that demonstrate the highest potential towards high PPW output solar cells. In this regard, we performed a Strength, Weaknesses, Opportunities and Threats (SWOT) analysis to evaluate OSCs and PSCs potential towards high PPW.

**Table 4.** SWOT Analysis of PSCs vs OSCs towards high PPW output.

| Solar technologies | Strength | Weakness |
|---|---|---|
| OSCs | <ul><li>Ease scaling up with R2R and other high-throughput printing fabrication techniques.</li><li>Low temperature processability</li><li>Suitable for flexible, lightweight applications</li><li>Extremely thin and light-weight materials used in the device stack significantly increase the devices' PPW output</li><li>Excellent mechanical flexibility passing several harsh bending and stretching tests</li></ul> | <ul><li>Relatively low conversion efficiency (up to 19.2% certified)</li><li>PCEs of OSCs are even lower when upscaled to full module size</li><li>PCE loss of approx. 4% when fabricated on ultrathin (1μm) substrates</li><li>Reliability of PCE may vary depending on the environmental factors during fabrication.</li><li>Limited lifespan of high PCE non-fullerene acceptor-based OSCs (up to 10 years of accelerated lifetime demonstrated)</li></ul> |
| PSCs | <ul><li>Scalable high throughput and R2R compatible PV technology</li><li>High PCE (competing with commercial silicon PVs) which highlights the potential towards high PPW outputs</li><li>Low formation energies for all deposited materials (thus</li></ul> | <ul><li>High PCE losses (approx. 5% when fabricated on ultrathin substrates)</li><li>Lead toxicity prevents PSCs to be applied on indoor, portable and wearable electronics</li></ul> |

| | | compatible with most ultrathin substrates)<br>• Long term stability of over 25 years with accelerated lifetime already demonstrated<br>• Extremely high tolerance towards high energy particles radiation<br>• Excellent mechanical flexibility passing several harsh bending and stretching tests | • Heavy metals such as lead or tin in the perovskite active layer significantly increases the overall device weight, thus reduce the PPW output<br>• Reproducibility issues at different environments during fabrication |
|---|---|---|---|
| **Solar technologies** | | **Opportunities** | **Threats** |
| **OSCs** | | • Great potential as power source in low-power consumption applications including indoor, IOT, portable and wearable electronics due to low toxicity materials<br>• Great potential towards EVs and outdoor power systems integration due to high record PPW outputs. | • Poor stability (of high PCE non-fullerene acceptor-based OSCs) under full illumination conditions may jeopardise their commercialization |
| **PSCs** | | • Great potential towards EVs, space applications and outdoor power systems integration due to high record PPW | • Perovskite active material toxicity poses significant environmental and health risks, thus proper encapsulation should be identified to prevent lead contamination.<br>• Material and panel recyclability |

Overall, both OSCs and PSCs are low-cost to produce compared to traditional inorganic solar PV technologies at high throughput, using solution processing techniques at low temperatures. Thus, they are fully compatible with ultra-thin and lightweight device manufacturing. They offer flexibility and versatility, through chemistry composition and interfacial engineering by modifying their absorbance spectrum accordingly to the application. OSCs and PSCs have unique strengths and weaknesses, and each technology has the potential to significantly impact the field of high PPW solar cells. As a conclusion, we can point out the competitive advantage of OSCs towards wearable and indoor application due to their eco-friendly character and of course their superior PPW values. On the other hand, the already proven long term stability of PSCs is giving them a competitive advantage towards exploiting them as power sources towards EVs and outdoor applications and their extremely high tolerance against high energy particles radiation is giving them a very good fit towards exploiting this PV technology for space applications. However, there is a lot of work still needs to be done for both high efficiency PSCs and OSCs to overcome the current challenges (mainly stability and reproducibility issues) they face before commercialization.

## 4.1 Potential and applications of solution-processed PVs with high PPW

After demonstrating our theoretical simulations and optimizations for OSCs and PSCs and comparing them with the current experimental results published in the literature, it is quite clear that both technologies are still far away from reaching their maximum potential regarding PPW outputs. If we focus only on the maximum theoretical potential, OSCs seem to be more promising towards achieving high PPW values. However, the poor operational stability of high efficiency non-fullerene acceptor-based OSCs and the very promising recent stability results for PSCs (several studies repot PSCs passing damp heat test and estimated accelerated lifetimes of more than 25 years), points out that there is still a lot of work needs to be done especially towards commercializing high PCE and PPW OSCs. If this stability issue cannot be solved, PSCs have a clear advantage especially for outdoor applications (EVs, outdoor IOT and space solar harvesting). Now regarding indoor applications, due to the absence of UV and IR and the significantly lower light intensity (there is no stability issue for OSC here) and the eco-friendly character of OSCs, they have a clear advantage compared to PSCs for exploiting high PPW figure of merit.

Despite the extremely high PPW potential values extracted from our simulations, there is a huge gap with the experimental results. On the one hand, for OSCs the record PPWs have been achieved with a device architecture utilizing materials and layers that most of them are providing high PPW but also some others than can be replaced by alternative ones. In particular, the 2.5 μm thick PET substrate is almost two-times thicker than the proposed PDMS substrate (PDMS has also much lower mass density) and the Ag metal electrode (this can be replaced by a much lighter Al metal electrode). These two suggested modifications, if can be applied without affecting the devices' PCE, a significant PPW output improvement is expected. On the other hand, for PSCs there also are a few parameters that should be improved in the state-of-the-art presented devices towards achieving ultra-high PPW values. These modifications include the replacement of PEN substrate (using the low mass density PDMS substrate, keeping its thickness same at 1.3μm), the utilization of thinner perovskite active layers (as thin as 350 nm) and with these conditions achieve significantly higher PCEs (much higher than the already presented ones of approx. 16%). The suggested modifications in both ultra-lightweight OSCs and PSCs are expected to at least double their PPW values even without further improving their PCEs (just because of the significant reduction of the devices' weight). Of course, to achieve the record high PPWs as presented in our simulations part, it is mandatory that these devices achieve remarkably high PCEs (such as 25% for PSCs and 20% for OSCs). This is not straightforward, and thus sophisticated methods for the controlled and uniform deposition and growth of all the layers used in the suggested device stacks must be developed and optimized. Furthermore, other open issues need to be solved to make these PV technologies even more competitive towards PPW figure of merit, include solving the stability issues of high-PCE non-fullerene acceptor- OSCs and improving the PCEs of ultrathin PSCs (16% for ultrathin compared to 26% for rigid PSCs).

Additionally, another point needs to be discussed here, include the large variety of potential applications these ultra-thin and light-weight PV technologies can be exploited. One of the most interesting applications of high PPW solar cells is in the field of solar-powered aircrafts, since reducing the weight of each part is maybe the most critical parameter. One way to achieve that is the use of ultra-lightweight high PPW solar cells. The low weight is a critical parameter here as it would contribute significantly to the reduction of energy demand for the flight. Furthermore, the solar panels should be efficient enough to supply energy to the propulsion system during the day while charging the battery units for night flights. Apart from solar-powered aircraft, high PPW solar cells could also benefit unmanned aerial vehicles (UAVs), drones and air balloons. Thus, it is obvious that both ultra-lightweight OSCs and PSCs could play a vital role for the development of solar power aircraft due to their groundbreaking potential towards high PPW outputs. Another interesting application for high PPW solar cells is wearable electronics since their integration with energy storage devices enables the self-powered systems to be completely independent and exempted from external charging. Nevertheless, until now most of the reported self-powered energy systems suffer from low conversion efficiency, and lack of flexibility and lightweight, which is far from meeting the requirements of wearable devices.[49] Among the various emerging PV technologies, OSCs have shown promising potential for integration in such applications due to their excellent PPW potential as well as their eco-friendly character (heavy-metal free solar cells). Internet-of-Things (IoT) technologies are alternative potential applications for high PPW solar cells, as this has been applied for the purpose of self-powered and more efficient energy usage and management.[50] Self-powered low-power consumption IoT platforms can be integrated either in indoor or outdoor systems depending on the purpose of the use. The use of OSCs could be implemented more effectively in indoor systems due to their greener character (PSCs contain toxic elements with the risk of contamination and human exposure) as well as their excellent PPW potential and stability against degradation (modern indoor lighting sources such as LEDs exclude the UV and IR of the light spectrum). On the other hand, PSCs seem to be more compatible in outdoor environments, because of their superior stability, performance under full sun illumination and their high PPW potential. However, solving the open research topic of improving the outdoor stability of high efficiency (non-fullerene acceptor-based) OSCs might put OSCs one step closer to their implementation in outdoor systems as well. Ultimately, the ability of OSCs and PSCs to persist for a long period of time without any need for human access or maintenance could facilitate the development of the next generation of energy-saving smart electric grids, powered by flexible and high PPW solar cells. The development of solar-powered electric vehicles (EVs) is another promising potential application of high PPW solar cells. Although EVs have zero greenhouse gasses emissions while running on electric power, the electricity that is required for charging the batteries is coming from the grid which contributes to $CO_2$ emissions.[51–53] Due to their high PCE values and their excellent stability in outdoor, PSCs are in the frontline to be integrated in EVs. The implementation of ultra-thin, lightweight, and high PPW PSCs on EVs, could provide extra driving range which could arise from their solar energy supply, significantly outperforming other commercially available EV integrated PV technologies (due to the superior PPWs). Here, another advantage of high PPW PSCs is their high degree of flexibility makes the installation of high PPW PSCs on the curved parts of the car's body more feasible while extending their driving range capabilities. Another appealing application of high PPW emerging solar cells is their integration into ultra-flexible optical-electronic devices, as energy-power sources. The so called, next generation of self-powered biomedical devices will need to be autonomous and bio-compatible to human skin or other tissues. Such device architectures would enable the accurate and continuous detection of physiological signals without the need for an external power supply or connecting wires.[54,55] Among the emerging the candidates of high PPW solar cells, OSCs embrace the absence of hazardous elements such as heavy metals, high degree of flexibility alongside with high PPW power output delivery which underlines the bright prospect of their employment in biomedical application concepts. Finally, solar cells have been demonstrated as the ideal choice as a power energy source for space applications. However, due to the lack of atmospheric scattering and absorption, the space environment consists of strong radiation (X-rays, gamma rays), affecting the performance until the complete failure of the electronic devices.[56] These high-energy particles are known to cause damage and degradation to semiconductors through ionization and displacement. As the tolerance of PSCs to proton irradiation has been explored, proving their long-term operation of high PPW PSCs in space could be the next step to serve as an ideal candidate among the emerging PV technologies in high-energy photon radiation conditions.[57] On top of that, PSCs combine high PPW output, strong high energy radiation absorption, flexibility, and low cost and thus, they hold the promise to serve as the next generation technology of solar-powered satellites as well as their implementation on the International Space Station and deep space solar-powered missions.[58,59]

## 4.2 Conclusions

In this perspective article, we explored the recent breakthroughs of the emerging solution-processed PV technologies (PSCs, OSCs and QDSCs) on flexible, ultra-light weight high PPW output concepts. The initial literature revealed that the most ideal and promising PV technologies are OSCs and PSCs mainly due to their ultra-low weight and combined high PCE respectively. For both OSCs and PSCs the maximum PPW values reported so far in the literature are approximately 30 W/g, outperforming by one order of magnitude the state-of-the art inorganic PV technologies (3.2 W/g for CIGS and a-Si). On the other hand, QDSCs reached PPW values of approximately 15 W/g (2 times lower than OSCs and PSCs) due to their low PCEs. To better understand the potential of OSCs and PSCs for high PPW, we performed a detailed theoretical simulation study, fully optimizing all the parameters (PCE and weight) towards high PPW outputs. Thus, meaningful insights regarding the dependency relationship between the PCE and weight with PPW output of PSCs and OSCs as a fraction to substrate thickness and mass density of the device were obtained. An impressive upper theoretical PPW limit of 74.3 and 93.7 W/g for PSCs and OSCs, respectively was calculated. Interestingly, different optimization approaches must be followed to achieve this, as suggested earlier in the theoretical analysis section of this work. As far as the current literature values are concerned, the highest reported values of both PV technologies are still lacking behind their maximum theoretical potential of PPW output. As it appears, PSCs have achieved until now only 40% of their theoretical PPW limit, whereas in the case of OSCs, the literature stands at almost 35%, respectively. Clearly, OSCs are considered to have a significant advantage towards the PSCs from a material perspective due to their significantly lower mass density, which shows their superiority and compatibility with high PPW PV applications. Furthermore, we have identified the key potential applications for high PPW solar cells, and we have sorted PSCs and OSCs according to their compatibility (when long term operation stabilities needed PSCs have a clear advantage, while when there is toxic element contamination issues, OSCs have a straightforward advantage) with these applications. Finally, we propose to the literature, that the calculation of the PPW output for the fabricated devices must take place within the natural dimensions of the active area of the cells. That way we avoid the overestimation of the PPW output of the device since the top metal electrode does not completely cover the area of the flexible ultra-thin solar cell (in this way it will be avoided the underestimation of the device's total weight and thus overestimation of PPW). Additionally, while up-scaling OSCs and PSCs it has been observed a significant PCE drop (approximately 1% for every decade increase in the active area). Therefore, we cannot calculate the PPW of a solar cell with a very small active area (<0.1 cm$^2$) considering it would have the same PCE even for large active areas (>1 cm$^2$).

## Conflicts of interest

There are no conflicts to declare.

## Acknowledgements

G.K. acknowledges the funding from the European Union's Horizon 2020 Research and Innovation program under the Marie Skłodowska-Curie grant agreement no. 101024237. E.A.A gratefully acknowledges the support of King Abdulaziz City for Science and Technology (KACST), Saudi Arabia. S.R.P.S., D.K. and A.P gratefully acknowledges the support of MUSICODE H2020, the European collaborative research and innovation project led by a multi-disciplinary consortium.

## Authors contributions

G.K. conceived the idea and supervised this work. G.K., A.P., T.M and T.D.A.. discussed and planned the content, analysed the literature, wrote the first draft. G.P and G.K. discussed and planned the simulations work. G.P. performed the theoretical simulations work and wrote the simulations part. E.A.A. contributed to the revision of the first draft. D.K. and F.H.I. contributed to the applications part. K.P., S.A., M.K., S.R.P.S., T.D.A. and M.G. contributed to the preparation of the manuscript. All authors contributed to the preparation of the final manuscript.

## Notes

The article has been submitted to Applied Physics Review. After it is published, it will be found at https://pubs.aip.org/aip/apr.

## References

[1] "Developments and Forecasts of Increasing Demographic Imbalances | Knowledge for policy," (n.d.).

[2] "BP forecasts robust global energy demand to 2030 despite efficiency gains | News and insights | Home," (n.d.).

[3] "EESI Environmental and Energy Study Institute eesi.org/topics/fossil-fuels/description," (n.d.).

[4] "BP Statistical Review of World Energy 2020 bp.com/content/dam/bp/business-sites/en/global/corporate/pdfs/energy-economics/statistical-review/bp-stats-review-2020-full-report," (n.d.).

[5] H. Damon Matthews, N.P. Gillett, P.A. Stott, and K. Zickfeld, "The proportionality of global warming to cumulative carbon emissions," Nature **459**, (2009).

[6] "Understanding global change ugc.berkeley.edu/background-content/burning-of-fossil-fuels," (n.d.).

[7] "UNFCCC. Adoption of the Paris Agreement FCCC/CP/2015/10/Add.1. (2015)," (n.d.).

[8] H.D. Matthews, and K. Caldeira, "Stabilizing climate requires near-zero emissions," Geophys. Res. Lett **35**, 4705 (2008).

[9] Joeri Rogelj, Michiel Schaeffer, Malte Meinshausen, Reto Knutti, Joseph Alcamo, Keywan Riahi, and William Hare, "Zero emission targets as long-term global goals for climate protection," Environ. Res. Lett **10**, 105007 (2015).

[10] B.M. Sanderson, B.C. O'neill, and C. Tebaldi, "Geophysical Research Letters What would it take to achieve the Paris temperature targets?," Geophys Res Lett **43**(13), 7133–7142 (2016).

[11] "World Nuclear Association Carbon Dioxide Emissions from Electricity https://www.world-nuclear.org/information-library/energy-and-the-environment/carbon-dioxide-emissions-from-electricity," (n.d.).

[12] "International Energy Agency Executive Summary Renewables 2021 iea.org/reports/renewables-2021/executive-summary," (n.d.).

[13] P. Reinhard, A. Chiriiii, P. Blasch, F. Pianezzi, S. Nishiwaki, S. Buecheler, and A.N. Tiwari, *Review of Progress Toward 20% Efficiency Flexible CIGS Solar Cells and Manufacturing Issues of Solar Modules* (2012).

[14] M.O. Reese, S. Glynn, M.D. Kempe, D.L. McGott, M.S. Dabney, T.M. Barnes, S. Booth, D. Feldman, and N.M. Haegel, "Increasing markets and decreasing package weight for high-specific-power photovoltaics," Nat Energy **3**, 1002–1012 (2018).

[15] M. Kaltenbrunner, M.S. White, E.D. Głowacki, T. Sekitani, T. Someya, and S. Bauer, "Ultrathin and lightweight organic solar cells with high flexibility," Nat Commun **3**, (2012).

[16] "NREL. Best Research Cell Efficiency Chart. https://www.nrel. gov/pv/assets/pdfs/best-research-cell-efficiencies," (n.d.).

[17] G. Li, R. Zhu, and Y. Yang, "Polymer solar cells," Nature Publishing Group **6**, 153–161 (2012).

[18] M.A. Green, A. Ho-Baillie, and H.J. Snaith, "The emergence of perovskite solar cells," Nature Publishing Group **8**, 506–514 (2014).

[19] M. Yang, D. Hoe Kim, T.R. Klein, Z. Li, M.O. Reese, B.J. Tremolet de Villers, J.J. Berry, M.F. A M van Hest, and K. Zhu, "Highly Efficient Perovskite Solar Modules by Scalable Fabrication and Interconnection Optimization," ACS Energy Lett **3**, 322–328 (2018).

[20] K. Fukuda, K. Yu, and T. Someya, "The Future of Flexible Organic Solar Cells," Adv Energy Mater **10**(25), 2000765 (2020).

[21] D. Du, F. Qiao, Y. Guo, F. Wang, L. Wang, C. Gao, D. Zhang, J. Liang, Z. Xu, W. Shen, and H. Wang, "Photovoltaic performance of flexible perovskite solar cells under bending state," Solar Energy **245**, 146–152 (2022).

[22] X. Ma, Q. Deng, L. Wang, X. Zheng, S. Wang, Q. Wang, L. Chen, L. Huang, X. Ouyang, and S. Cao, "Cellulose transparent conductive film and its feasible use in perovskite solar cells," RSC Adv **9**(17), 9348–9353 (2019).

[23] N.S., P.H.E., H.H.Q.& S.P.R. Fatemi, "Solar Arracy Trades between very High-Efficiency Multi-Junction and Si Space Solar Cells. in IEEE PVSC 15–18 (2000).," Conference Record of the Twenty-Eighth IEEE Photovoltaic Specialists Conference, (2000).

[24] S. Kang, J. Jeong, S. Cho, Y.J. Yoon, S. Park, S. Lim, J.Y. Kim, and H. Ko, "Ultrathin, lightweight and flexible perovskite solar cells with an excellent power-per-weight performance," J Mater Chem A Mater **7**(3), 1107–1114 (2019).

[25] W. Song, K. Yu, E. Zhou, L. Xie, L. Hong, J. Ge, J. Zhang, X. Zhang, R. Peng, and Z. Ge, "Crumple Durable Ultraflexible Organic Solar Cells with an Excellent Power-per-Weight Performance," Adv Funct Mater **31**(30), 2102694 (2021).

[26] M. Kaltenbrunner, G. Adam, E.D. Głowacki, M. Drack, R. Schwödiauer, L. Leonat, D.H. Apaydin, H. Groiss, M.C. Scharber, M. Schuette White, N.S. Sariftci, and S. Bauer, "Flexible high power-per-weight perovskite solar cells with chromium oxide-metal contacts for improved stability in air," (2015).

[27] M. Li, J. Zhou, L. Tan, H. Li, Y. Liu, C. Jiang, Y. Ye, L. Ding, W. Tress, and C. Yi, "Multifunctional succinate additive for flexible perovskite solar cells with more than 23% power-conversion efficiency," Innovation **3**(6), (2022).

[28] Z. Liu, P. You, C. Xie, G. Tang, and F. Yan, "Ultrathin and flexible perovskite solar cells with graphene transparent electrodes," Nano Energy **28**, 151–157 (2016).

[29] H. Zhang, J. Cheng, D. Li, F. Lin, J. Mao, C. Liang, A.K.Y. Jen, M. Grätzel, and W.C.H. Choy, "Toward All Room-Temperature, Solution-Processed, High-Performance Planar Perovskite Solar Cells: A New Scheme of Pyridine-Promoted Perovskite Formation," Advanced Materials **29**(13), 1604695 (2017).

[30] Q. Wang, C.-C. Chueh, T. Zhao, J. Cheng, M. slamian, W.C. H Choy, and A. K-YJ en, "Effects of Self-Assembled Monolayer Modification of Nickel Oxide Nanoparticles Layer on the Performance and Application of Inverted Perovskite Solar Cells," (n.d.).

[31] M. Xie, J. Wang, J. Kang, L. Zhang, X. Sun, K. Han, Q. Luo, J. Lin, L. Shi, and C.Q. Ma, "Super-flexible perovskite solar cells with high power-per-weight on 17 μm thick PET substrate utilizing printed Ag nanowires bottom and top electrodes," Flexible and Printed Electronics **4**(3), 034002 (2019).

[32] H. Li, X. Li, W. Wang, J. Huang, J. Li, Y. Lu, J. Chang, J. Fang, and W. Song, "Highly Foldable and Efficient Paper-Based Perovskite Solar Cells," Solar RRL **3**(3), 1800317 (2019).

[33] J. Jean, A. Wang, and V. Bulović, "In situ vapor-deposited parylene substrates for ultra-thin, lightweight organic solar cells," Org Electron **31**, 120–126 (2016).

[34] Z. Jiang, F. Wang, K. Fukuda, A. Karki, W. Huang, K. Yu, T. Yokota, K. Tajima, T.Q. Nguyen, and T. Someya, "Highly efficient organic photovoltaics with enhanced stability through the formation of doping-induced stable interfaces," Proc Natl Acad Sci U S A **117**(12), 6391–6397 (2020).

[35] E. Bihar, D. Corzo, T.C. Hidalgo, D. Rosas-Villalva, K.N. Salama, S. Inal, and D. Baran, "Fully Inkjet-Printed, Ultrathin and Conformable Organic Photovoltaics as Power Source Based on Cross-Linked PEDOT:PSS Electrodes," Adv Mater Technol **5**(8), 2000226 (2020).

[36] J. Wan, R. Wen, Y. Xia, M. Dai, H. Huang, L. Xue, Z. Zhang, J. Fang, K.N. Hui, and X. Fan, "All annealing-free solution-processed highly flexible organic solar cells," J Mater Chem A Mater **9**(9), 5425–5433 (2021).

[37] X. Zhang, V.A. Öberg, J. Du, J. Liu, and E.M.J. Johansson, "Extremely lightweight and ultra-flexible infrared light-converting quantum dot solar cells with high power-per-weight output using a solution-processed bending durable silver nanowire-based electrode," Energy Environ Sci **11**(2), 354–364 (2018).

[38] M.M. Tavakoli, M.H. Gharahcheshmeh, N. Moody, M.G. Bawendi, K.K. Gleason, and J. Kong, "Efficient, Flexible, and Ultra-Lightweight Inverted PbS Quantum Dots Solar Cells on All-CVD-Growth of Parylene/Graphene/oCVD PEDOT Substrate with High Power-per-Weight," Adv Mater Interfaces **7**(16), 2000498 (2020).

[39] W. Shockley, and H.J. Queisser, "Detailed Balance Limit of Efficiency of p-n Junction Solar Cells Detailed Balance Limit of Efficiency of p-n Junction Solar Cells*," J Appl Phys **32**, 510–519 (1961).

[40] L. Zhu, M. Zhang, J. Xu, C. Li, J. Yan, G. Zhou, W. Zhong, T. Hao, J. Song, X. Xue, Z. Zhou, R. Zeng, H. Zhu, C.C. Chen, R.C.I. MacKenzie, Y. Zou, J. Nelson, Y. Zhang, Y. Sun, and F. Liu, "Single-junction organic solar cells with over 19% efficiency enabled by a refined double-fibril network morphology," Nature Materials 2022 21:6 **21**(6), 656–663 (2022).

[41] W.A. Laban, and L. Etgar, "Depleted hole conductor-free lead halide iodide heterojunction solar cells," Energy Environ Sci **6**(11), 3249–3253 (2013).

[42] G. Perrakis, A.C. Tasolamprou, G. Kenanakis, E.N. Economou, S. Tzortzakis, and M. Kafesaki, "Submicron Organic-Inorganic Hybrid Radiative Cooling Coatings for Stable, Ultrathin, and Lightweight Solar Cells," ACS Photonics **9**(4), 1327–1337 (2022).

[43] "'Material Science | News | Materials Engineering | News,' https://www.azom.com/.," (n.d.).

[44] Y. Shi, W. Li, A. Raman, and S. Fan, "Optimization of Multilayer Optical Films with a Memetic Algorithm and Mixed Integer Programming," ACS Photonics **5**(3), 684–691 (2018).

[45] G. Khanarian, J. Joo, X. Liu, J. Chem Phys, P. Eastman, D. Werner, and P. Trefonas, "The optical and electrical properties of silver nanowire mesh films," J Appl Phys **114**, 24302 (2013).

[46] Z. Li, B. Li, X. Wu, S.A. Sheppard, S. Zhang, D. Gao, N.J. Long, and Z. Zhu, "Organometallic-functionalized interfaces for highly efficient inverted perovskite solar cells," Science (1979) **376**(6591), (2022).

[47] E.A. Alharbi, A. Krishna, N. Lempesis, M. Dankl, I. Mosquera-Lois, M.A. Hope, T.P. Baumeler, G. Kakavelakis, A. Mishra, F.T. Eickemeyer, O. Ouellette, T. Chawanpunyawat, A. Hagfeldt, S.M. Zakeeruddin, L. Emsley, L. Pfeifer, U. Roethlisberger, and M. Grätzel, "Cooperative passivation of perovskite solar cells by alkyldimethylammonium halide amphiphiles," Joule **7**(1), 183–200 (2023).

[48] T.P. Baumeler, E.A. Alharbi, G. Kakavelakis, G.C. Fish, M.T. Aldosari, M.S. Albishi, L. Pfeifer, B.I. Carlsen, J.H. Yum, A.S. Alharbi, M.D. Mensi, J. Gao, F.T. Eickemeyer, K. Sivula, J.E. Moser, S.M. Zakeeruddin, and M. Grätzel, "Surface Passivation of FAPbI3-Rich Perovskite with Cesium Iodide Outperforms Bulk Incorporation," ACS Energy Lett, 2456–2462 (2023).

[49] J. Zhao, Z. Xu, Z. Zhou, S. Xi, Y. Xia, Q. Zhang, L. Huang, L. Mei, Y. Jiang, J. Gao, Z. Zeng, and C. Tan, "A Safe Flexible Self-Powered Wristband System by Integrating Defective $MnO_{2-x}$ Nanosheet-Based Zinc-Ion Batteries with Perovskite Solar Cells," ACS Nano **15**(6), 10597–10608 (2021).

[50] Q. Chang, D. Wu, Y. Huang, C. Liang, L. Liu, H. Liu, Y. Liu, J. Qiu, X. Tang, and G. Han, "Ultrahigh Response Humidity Sensor Based on Lead-Free $Cs_2SnCl_6$ Perovskite Films," IEEE Electron Device Letters **43**(5), 805–808 (2022).

[51] G.R. Chandra Mouli, P. Bauer, and M. Zeman, "System design for a solar powered electric vehicle charging station for workplaces," Appl Energy **168**, 434–443 (2016).

[52] F. Ahmad, M. Khalid, and B.K. Panigrahi, "An enhanced approach to optimally place the solar powered electric vehicle charging station in distribution network," J Energy Storage **42**, 103090 (2021).

[53] S. Lee, S. Iyengar, D. Irwin, and P. Shenoy, "Shared solar-powered EV charging stations: Feasibility and benefits," 2016 7th International Green and Sustainable Computing Conference, IGSC 2016, (2017).

[54] A. Joseph, A.B. Pillai, V.K. Pulikodan, A. Alexander, R. Muhammed, and M.A.G. Namboothiry, "Solution-Processed Self-Powered Panchromatic Organic Photodiode and Its Application in Biomedical Devices," ACS Appl Electron Mater **4**(4), 1567–1575 (2022).

[55] S. Park, S. Won Heo, W. Lee, D. Inoue, Z. Jiang, K. Yu, H. Jinno, D. Hashizume, M. Sekino, tomoyuki Yokota, K. Fukuda, K. tajima, and takao Someya, "Self-powered ultra-flexible electronics via nano-grating-patterned organic photovoltaics," (2018).

[56] Y. Tu, J. Wu, G. Xu, X. Yang, R. Cai, Q. Gong, R. Zhu, and W. Huang, "Perovskite Solar Cells for Space Applications: Progress and Challenges," Advanced Materials **33**(21), 2006545 (2021).

[57] G. Kakavelakis, M. Gedda, A. Panagiotopoulos, E. Kymakis, T.D. Anthopoulos, and K. Petridis, "Metal Halide Perovskites for High-Energy Radiation Detection," Advanced Science **7**(22), 2002098 (2020).

[58] F. Lang, M. Jošt, J. Bundesmann, A. Denker, S. Albrecht, G. Landi, H.C. Neitzert, J. Rappich, and N.H. Nickel, "Efficient minority carrier detrapping mediating the radiation hardness of triple-cation perovskite solar cells under proton irradiation," Energy Environ Sci **12**(5), 1634–1647 (2019).

[59] W. Delmas, S. Erickson, J. Arteaga, M. Woodall, M. Scheibner, T.S. Krause, K. Crowley, K.T. Vansant, J.M. Luther, J.N. Williams, J. Mcnatt, T.J. Peshek, L. Mcmillon-Brown, S. Ghosh, W. Delmas, S. Erickson, J. Arteaga, M. Woodall, M. Scheibner, S. Ghosh, and T.S. Krause, "Evaluation of Hybrid Perovskite Prototypes After 10-Month Space Flight on the International Space Station," Adv Energy Mater **13**(19), 2203920 (2023).